\documentclass[fleqn,10pt,table]{wlscirep}
\usepackage[utf8]{inputenc}
\usepackage[T1]{fontenc}
\usepackage{booktabs}
\usepackage{siunitx}
\usepackage{makecell}
\usepackage{textcomp}
\usepackage{multirow}
\usepackage{etoolbox,siunitx}
\usepackage{listings}
\usepackage{lineno}

\usepackage[disable]{todonotes}
\usepackage{graphicx}
\usepackage{lipsum}
\usepackage{subfig}

\robustify\bfseries

\definecolor{lightgray}{gray}{0.9}
 
\title{Machine learning on DNA-encoded libraries: A new paradigm for hit-finding}

\author[1,+]{Kevin McCloskey}
\author[2,3,+]{Eric A. Sigel}
\author[1]{Steven Kearnes}
\author[2]{Ling Xue}
\author[2]{Xia Tian}
\author[2,4]{Dennis Moccia}
\author[2]{Diana Gikunju}
\author[2]{Sana Bazzaz}
\author[2]{Betty Chan}
\author[2]{Matthew A. Clark}
\author[2,3]{John W. Cuozzo}
\author[2]{Marie-Aude Gui\'{e}}
\author[2]{John P. Guilinger}
\author[2,3]{Christelle Huguet}
\author[2]{Christopher D. Hupp}
\author[2]{Anthony D. Keefe}
\author[2,3]{Christopher J. Mulhern}
\author[2]{Ying Zhang}
\author[1,*]{Patrick Riley}
\affil[1]{Google Research Applied Science, Mountain View, CA, USA}
\affil[2]{X-Chem, Waltham, MA, USA}
\affil[3]{ZebiAI, Waltham, MA, USA}
\affil[4]{Cognitive Dataworks, Amesbury, MA , USA}
\affil[+]{Contributed equally to this work.}
\affil[*]{pfr@google.com}

\begin{abstract}
DNA-encoded small molecule libraries (DELs) have enabled discovery of novel inhibitors for many distinct protein targets of therapeutic value through screening of libraries with up to billions of unique small molecules.
We demonstrate a new approach applying machine learning to DEL selection data by identifying active molecules from a large commercial collection and a virtual library of easily synthesizable compounds. We train models using only DEL selection data and apply automated or automatable filters with chemist review restricted to the removal of molecules with potential for instability or reactivity.
We validate this approach with a large prospective study (nearly 2000 compounds tested) across three diverse protein targets: sEH (a hydrolase), ER$\alpha$ (a nuclear receptor), and c-KIT (a kinase).
The approach is effective, with an overall hit rate of ${\sim}$30\% at 30~{\textmu}M and discovery of potent compounds (IC$_{50}$ <10~nM) for every target.
The model makes useful predictions even for molecules dissimilar to the original DEL and the compounds identified are diverse, predominantly drug-like, and different from known ligands.
Collectively, the quality and quantity of DEL selection data; the power of modern machine learning methods; and access to large, inexpensive, commercially-available libraries creates a powerful new approach for hit finding.

\end{abstract}
\begin{document}

\flushbottom
\maketitle
\thispagestyle{empty}

\listoftodos


\section*{Introduction}
Discovering small molecule therapeutics is an increasingly expensive and long process\cite{DiMasi2016-nl}. Once a target is validated, finding diverse small molecule hits that modulate its function is foundational for a successful drug discovery effort. 
These hits should also have good physicochemical properties and be tractable for further optimization into therapeutic candidates. 
Effective computational screening of large virtual libraries has long been a goal of the community. Here, we present a new process for building a machine learned model from readily generated experimental data and using that model on large, low-cost chemical libraries. 
We validate this approach with the largest reported prospective experimental study using machine learning (ML) for hit finding.

DNA encoded small molecule libraries (DELs)\cite{Clark2009-uz} have been increasingly explored in recent years to enhance hit identification efforts in drug discovery. Capitalizing on the power of next generation sequencing (NGS) and reduced cost per compound tested as compared to high-throughput screening (HTS), this approach allows simultaneous readout of target binding by millions to billions of molecules\cite{Clark2009-uz,Goodnow2017DNAencoded,Harris2016DNAenco}. 
Accordingly, the use of DEL screening has significantly expanded the accessible scope of chemical space that can be explored in a single experiment, in terms of diversity and degree of variation around structural motifs \cite{Goodnow2017DNAencoded}. Success using DELs has been demonstrated across a broad range of targets of varied classes\cite{machutta2017prioritizing} by multiple pharmaceutical, biotech, and academic groups\cite{Goodnow2017DNAencoded}. A number of programs based on DEL-identified hits have progressed to clinical trials\cite{Harris2017-ck,Belyanskaya2017-mn}.

However, existing successes have limitations. Analysis of DEL selections has typically focused on identifying molecules within the DEL by directly examining the output, aided by informatics analysis and visualization tools \cite{Clark2009-uz,Goodnow2017DNAencoded}. This close involvement of human analysis limits the scale of molecules considered, introduces bias, and makes it difficult to fully utilize the subtle patterns in the DEL selections. These subtle patterns may be obscured by sources of variability such as the yield of individual library members and random sampling effects.\cite{Satz2017-zh,Kuai2018-ua}.

Over the last decade, neural networks have demonstrated strong performance on molecular property prediction tasks\cite{gilmer2017neural,Smith2017-tx,Schutt2018-jo,Li2019-ha,Wu2018-dp,Lenselink2017-ve}. 
For many applications in drug discovery with small and/or sparse data, neural network methods do not outperform simpler methods like random forests\cite{Wu2018-dp,kearnes2016molecular}; however, the benefits of custom graph-based architectures become clear with large\cite{ma2015deep} or highly structured\cite{gilmer2017neural} data, and DEL selection data is both.

\begin{figure}[htb]
\centering
\includegraphics[width=\linewidth]{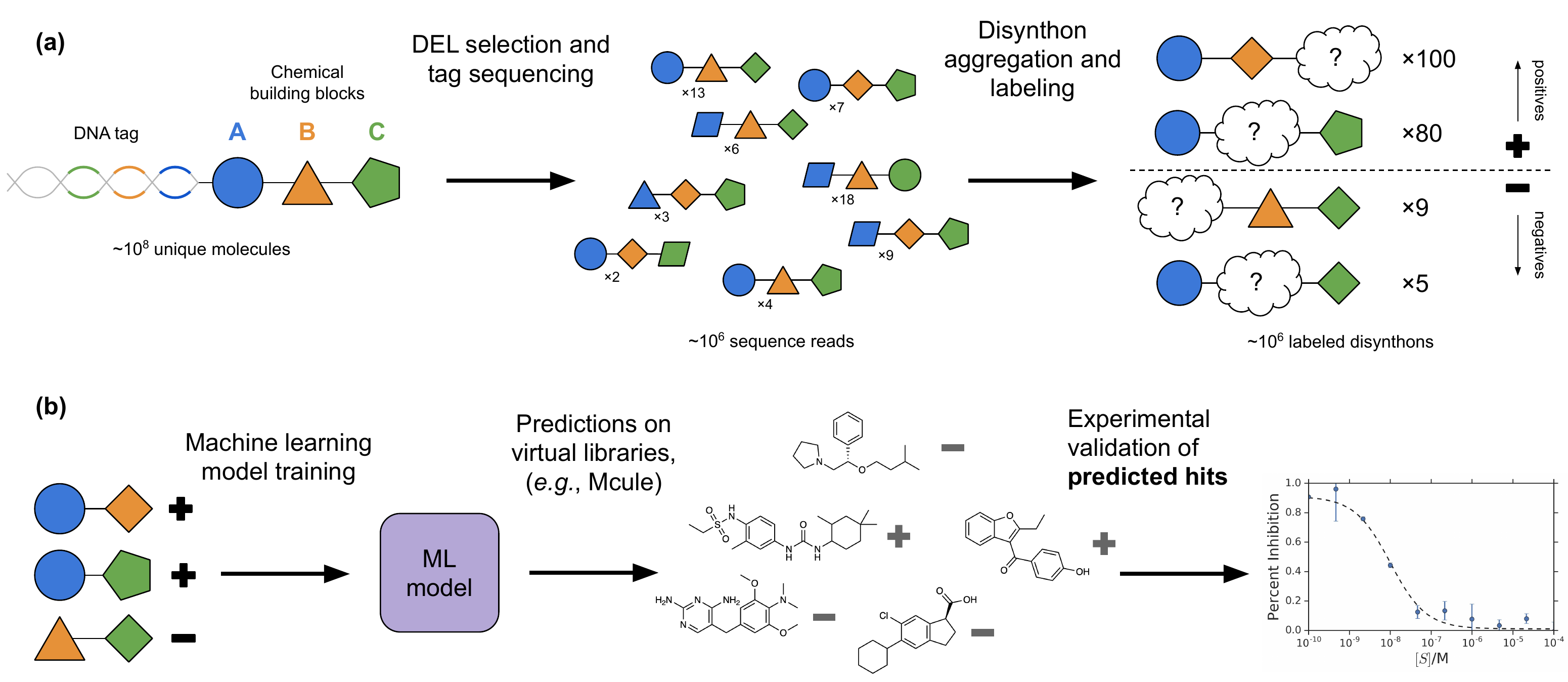}
\caption{Schematic example of machine learning models trained on DEL data. \textbf{(a)} Starting with a DEL containing ${\sim}10^{8}$ unique molecules, an affinity-mediated selection is performed against the target and the DNA tags for retained molecules are PCR-amplified and sequenced. After removal of PCR-amplification duplicates, reads for each library member are then aggregated across shared two-cycle disynthon representations. These disynthons are labeled as positive or negative based on calculated enrichment scores. Aggregation is performed for every possible pair of synthons; \emph{i.e.}, some disynthons aggregate over the central synthon(s). The figure shows an example for a three-cycle DEL, but we also used two-cycle and four-cycle libraries; overall, we ran selections for ${\sim}40$ libraries covering ${\sim}10^{11}$ unique molecules. Note that additional counter-selections may be run to provide richer labels, \emph{e.g.}, inclusion of a known competitive inhibitor. \textbf{(b)} The labeled disynthon representations are used as training data for machine learning models. The trained models are then used to predict hits from virtual libraries or commercially available catalogs such as Mcule. Predicted hit compounds are ordered or synthesized and tested experimentally to confirm activity in functional assays.}
\label{fig:scheme}
\end{figure}

In this work, we demonstrate a new application of DEL selection data for discovering hits outside the compounds in the DEL (\figurename~\ref{fig:scheme}).
First, affinity-mediated selections of the DEL under several conditions were performed with each target.
Second, the sequencing readout was processed and aggregated (see Methods). 
Third, a machine learning model was trained on the aggregated selection data (using no prior off-DNA activity measurements) and used to virtually screen large libraries (${\sim}$88~M) of easily synthesizable or inexpensive purchasable compounds. 
Fourth, automated diversity filters, reactive substructure filters, and a chemist review restricted to elimination of molecules with potential instability or reactivity were applied to the top predictions of the model. 
Finally, the selected compounds were tested experimentally.

We show that graph convolutional neural network (GCNN) models\cite{kearnes2016molecular} trained with this approach generalize well to new chemical spaces and have much stronger prospective performance than simpler baseline models. For GCNN models applied to three different protein targets, we report hit rates for the best-performing target of 72\% at 30~{\textmu}M, 33\% at 10~{\textmu}M, and 29\% at 1~{\textmu}M. This is in contrast to traditional HTS (without ML), which normally reports hit rates of ${\sim}$1\%\cite{bender2008aspects,  clare2019industrial}. Our results demonstrate that this approach significantly expands the utility of DEL selection data by identifying hits in low-cost compound libraries, producing structurally diverse starting points for both tool compound discovery and lead generation at a fraction ($\sim25\%$) of the cost of typical DEL-based hit finding.

\section*{Results}


\subsection*{Discovering potent ligands}

Three therapeutic protein targets were screened: Soluble Epoxide Hydrolase (sEH) is a target for cardiovascular diseases\cite{imig2009soluble}, Tyrosine-protein kinase KIT (c-KIT) is a target for multiple pathologies including gastrointestinal stromal tumors\cite{rubin2007gastrointestinal}, and Estrogen Receptor Alpha (ER$\alpha$) is a target for multiple pathologies including breast cancer\cite{thomas2011different}.

Two types of ML models were trained on the DEL selection data to classify compounds: Random Forest (RF) \cite{breiman2001rf} and GCNN\cite{kearnes2016molecular}. The training data were preprocessed with disynthon aggregation (see Methods) to handle noise in DNA-sequencing counts of individual library members, e.g. due to undersampling of the DEL selection output (see \figurename~\ref{fig:scheme}). Notably, only the DEL selection data and ML techniques described herein were used in building these models---no known ligand data were used beyond the choice of the competitive inhibitors used in the DEL selections, and no explicit representation of the protein targets nor 3D data were used. In fact, the authors building the GCNN models were intentionally blinded to the names and nature of the targets at the time of model building. To cleanly assess the quality of the model predictions, we avoided subjective selection of the most chemically attractive compounds from the predictions. To identify molecules for purchase and testing, we started with the top predicted molecules and applied diversity, logistical, structural filters and a restricted chemist review (see Methods). Though not automated in this experiment, this limited chemist review could be automated.
All compounds successfully acquired or synthesized were experimentally validated.

Performance of an ML model is dependent on the data set it is trained on. In a traditional DEL screening approach, a single selection campaign is generally sufficient for hit identification against the target of interest. To ensure this is equally true for training predictive models, two separate DEL selections were performed months apart on sEH. This experiment showed that the two separate training sets were equivalent with respect to model training (see Methods and Extended Data \figurename~\ref{fig:two_selections_roc_curve}).

\begin{figure}[htb]
\centering
\includegraphics[width=\linewidth]{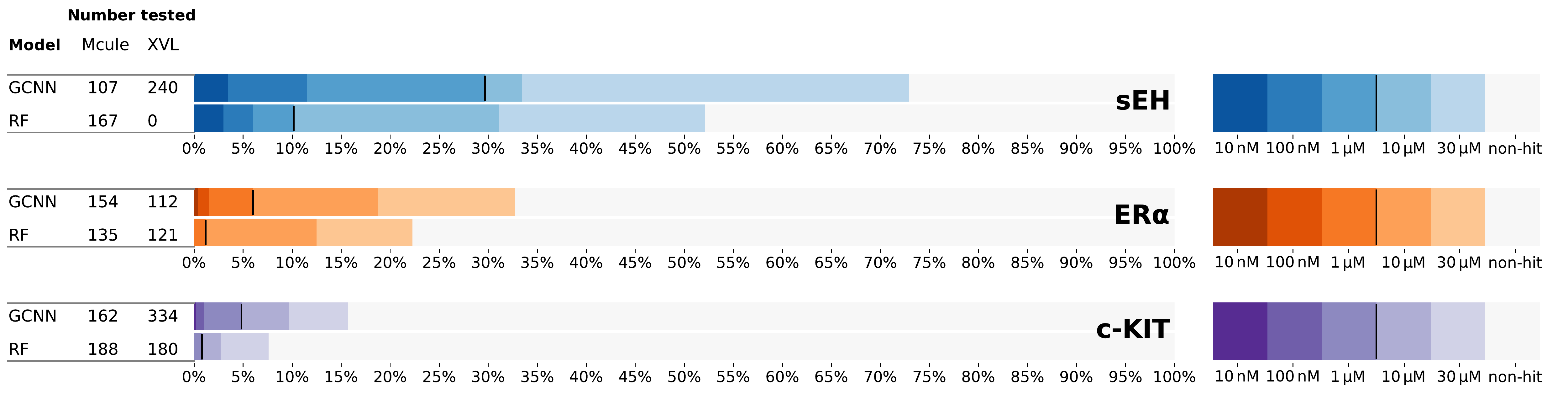}
\caption{Numbers tested along with hit rates and potencies across three therapeutic protein targets for two machine-learning models. Compounds came from Mcule, a commercial provider, and a proprietary virtual library (XVL). Lower concentrations correspond to more potent hits and are represented by darker colors; a black vertical line marks the 1~{\textmu}M threshold in each bar chart. Note that some compounds appeared in multiple target/model (\emph{e.g.}, ``sEH/GCNN'') buckets, such that the number of unique molecules is slightly smaller than the sum of the counts shown here (1885 vs. 1900).
}
\label{fig:hit_rates}
\end{figure}

Experimental validation followed a traditional two step approach: single-point inhibition assays were run first, followed by dose--response assays to confirm hits from the initial assays (see Methods). Dose--response potency values are reported as the concentration required for 50\% inhibition (IC$_{50}$). The experimental hit rates and potencies are reported in \figurename~\ref{fig:hit_rates} and cover 1885 unique compounds from two readily accessible, low-cost libraries: Mcule\cite{kiss2012http} and a proprietary single reaction virtual library (XVL; see Methods). Results from these two screening libraries have been combined for the main figures in this article. Notably, all the experimental validations in this work are biochemical activity or ligand displacement assays, reducing the likelihood of false positive hits that are inactive (non-binders, allosteric binders or silent binders).

Across the three protein targets, we identified 304 ligands with better than 10~{\textmu}M potency, and 165 with better than 1~{\textmu}M potency. The GCNN models achieved substantially higher hit rates and better potencies than the RF models. While the hit rates varied across protein targets, the GCNN model still identified 78 hits <30~{\textmu}M for the least productive protein (c-KIT). Hit rates may be correlated with the number of positive training examples (see Methods): sEH models had the highest hit rates and largest number of positive training examples, while c-KIT models had the lowest hit rates and fewest positive training examples. The Mcule library was generally more productive than the virtual library in terms of potency and hit rates (Extended Data \figurename~\ref{fig:hit_rates_mcule_and_vl1}). 
Perhaps not surprisingly---considering our filtering criteria and that Mcule and XVL are curated to be more drug-like---568/583 (97\%) of the unique confirmed hits had $\le 1$ Lipinski "Rule of 5" violations\cite{lipinski1997experimental} (Extended Data \figurename~\ref{fig:lipinski}). Some structures may still look unattractive to a skilled chemist; this is a result of our desire to limit subjective intervention.

As a baseline comparison, we also tested 107 compounds identified by a similarity search against a subset of positive training examples from the ER$\alpha$ DEL selection that were chosen for both high enrichment and diversity (see Methods). This similarity search yielded no hits with detectable activity. Because this approach found zero hits, we did not repeat this baseline for the other targets.

\subsection*{Analysis of confirmed hits discovered by ML}

As drug discovery campaigns move from hit-finding into lead optimization, the structural diversity of the hits matters: diverse hits act as insurance against local minima in the multi-objective lead optimization landscape \cite{bleicher2003guide}. Despite its large size (up to ${\sim}10^{11}$ molecules), a DEL represents a minute fraction of the universe of small, drug-like molecules (estimated at $10^{33}$ molecules\cite{polishchuk2013estimation}), so the degree to which the ML model is accurate far from the training data is paramount. Yet---across many applications---ML models often fail to generalize 
when tested on data distributions different from the training data\cite{zadrozny2004learning, sugiyama2005input}.

The development of simple metrics to evaluate similarity and diversity of small molecules remains an unsolved cheminformatics problem. No single metric has captured all the nuances, including differences in molecular size and domain/target-specific knowledge of what substitutions have similar effects. The most commonly used metric is Tanimoto similarity on Extended-Connectivity Fingerprints\cite{rogers2010extended} (ECFP) and their ``functional class'' counterpart (FCFP), see Methods for details. Another way to analyze similarity is with Bemis--Murcko scaffolds\cite{bemis1996properties}, which define a central structure that can be decorated with functional groups.

\begin{figure}[htb]
\centering
\includegraphics[width=\linewidth]{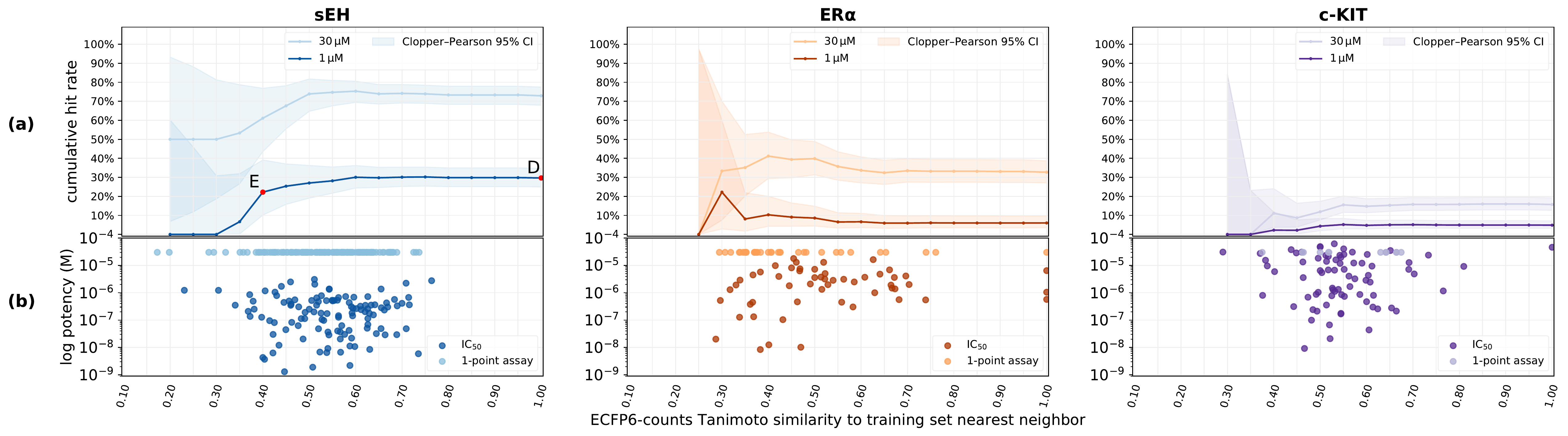}
\caption{Cumulative hit rates of GCNN-predicted compounds \textbf{(a)}, along with a scatter plot of hits \textbf{(b)}, on a shared x-axis of ECFP6-counts Tanimoto similarity of compounds to the training DELs. The cumulative hit rate plots show the hit rates for compounds with $\leq$ a given (x-axis) similarity to the training set. For example, the observed sEH hit rate at 1~{\textmu}M was 29.7\% (point \textbf{D} for sEH, 347 compounds tested), but when only considering compounds that have $\leq$~0.40 similarity to the training set nearest neighbor (point \textbf{E}, 36 compounds tested), the hit rate drops to 22.2\%. Error bands are Clopper--Pearson intervals\cite{clopper1934use} at 95\% confidence.}
\label{fig:hit_rate_by_sim}
\end{figure}

\figurename~\ref{fig:hit_rate_by_sim} depicts the cumulative hit rate and potency as a function of similarity to the nearest neighbor in the training set. While there is evidence of a drop off in hit rate as compounds become dissimilar from the training data, the hit rates remain useful even at less than 0.4 ECFP Tanimoto similarity to the training set (22, 28, and 5 hits with better than 30~{\textmu}M potency for sEH, ER$\alpha$ and c-KIT respectively); this suggests that GCNN models have the ability to generalize to unseen regions of chemical space. Many potent hits were found far from the training set (\emph{e.g.}, the hit least similar to ER$\alpha$ training data---with ECFP Tanimoto similarity of only 0.29 to the training set---had an IC$_{50}$ of 20~nM). Extended Data \figurename~\ref{fig:hit_rate_by_sim_extended} includes similar analysis with FCFP and produces comparable conclusions about generalization far from the DEL. Overall, there was no meaningful correlation between the biochemical IC$_{50}$ of identified hits and ECFP Tanimoto similarity to the DEL selection training set: the largest $R^2$ (squared regression correlation coefficient) on any target for GCNN predicted hits and RF predicted hits were 0.001 and 0.183 respectively. Extended Data \figurename~\ref{fig:hits_neighbors} shows distributions of similarity between confirmed hits and nearest training set compounds, while Extended Data Table~\ref{table:hits} highlights a selection of hits along with their nearest neighbors in the training set (to ground these similarity numbers with specific examples). Of the Bemis--Murcko scaffolds found in the confirmed hits, only 42.7\% (GCNN) and 60.8\% (RF) were also contained in the training set.

We applied diversity filtering (see Methods) in selecting compounds for testing. The final hits maintain diversity, as illustrated by Extended Data \figurename~\ref{fig:hit_vs_hit}(b) and scaffold analysis: the 418 hits with $\le$30~{\textmu}M potency identified from GCNN predictions were distributed among 370 unique Bemis--Murcko scaffolds, while the 170 hits identified from RF predictions were distributed among 166 scaffolds.

The confirmed hits are also structurally novel: only 2.2\% (GCNN) and 3.0\% (RF) of hit scaffolds were previously reported in ChEMBL\cite{gaulton2016chembl} for these targets and Extended Data \figurename~\ref{fig:hit_vs_hit}(b) shows distributions of similarity between confirmed GCNN hits and the nearest ChEMBL ligand (see Methods).

\section*{Discussion}

Overall, we have demonstrated a new virtual screening approach that couples DEL selection data with machine learning and automated or automatable filters to discover diverse, novel hits outside the DEL.
 This approach is effective on three diverse protein targets.
Because of the generalization of the ML models, practitioners have significant power in choosing a virtual library. They could restrict screening to molecules with desirable properties, such as synthesizability, commercial availability, presence of favored substructures, specific molecule property ranges, or dissimilarity to known ligands.
In this work, we focused on purchasable or easily-synthesizable molecules which tended to have drug-like properties. 
This avoids the time-consuming and expensive process of building new chemical matter into a DEL library and performing new selections, or incorporating new molecules into an HTS screening library. 
This ability to consider compounds outside of the DEL is the biggest advantage of our approach; notably, this approach can be used at a fraction of the cost of a traditional DEL screening followup, driven primarily by the large difference in synthesis cost (see Extended Data Table~\ref{table:del_vs_delml}).

The success of this approach is attributable to at least three factors: First, the past few years have seen the rise of more powerful machine learning methods for many problems. For hit-finding in particular, we provide the first large scale prospective evidence of modern graph based neural networks having a significant advantage over simpler methods. Second, DEL selection generates both the large quantity and the high quality of data points that is essential for the training of performant machine learning models. Lastly, large make-on-demand small molecule libraries (proprietary or commercially available) provide a source of low-cost, structurally diverse compounds for virtual screening. Just as Lyu~\emph{et al.}~\cite{Lyu2019-dg} showed effective use of commercially available libraries for a computational molecular docking screen, we have shown the utility of these libraries for machine learning driven screens.

We believe the ability of a model trained on binding data to predict activity comes in part from classification criteria that include DEL selection with a competitive binder (which may or may not be a small molecule) present in the target active site of interest. 
Future application of this approach could explore areas complementary to traditional HTS (as non-ML virtual screening has\cite{ferreira2010complementarity}), as well as integration with lead generation and optimization in combination with machine driven exploration of chemical space (such as Zhavoronkov \emph{et al.}\cite{zhavoronkov2019ddr1}).
We expect the impact of this approach to expand as DEL selections are used to measure properties beyond competitive on-target binding; for example, some absorption, distribution, metabolism, excretion, and toxicity (ADMET) properties may be assayable as DEL affinity-mediated screens.

The trends of more powerful machine learning and larger, more diverse make-on-demand libraries will continue, suggesting that the utility of the approach demonstrated here will grow over time. Further, with growth in the quality of the models and the number of targets to which they are applied, we hope to impact later stages of the drug discovery process.

\section*{Methods}
\label{sec:methods}

\subsection*{Machine learning and cheminformatics}

\subsubsection*{Classification of Disynthons from Selection Data} \label{methods:classes}

Sequencing data for each selection condition was compiled as summed counts for all combinations of two building blocks across all cycle combinations. For example, for a three cycle library of the form A-B-C, sums aggregating counts for the A-B, A-C, and B-C disynthons were generated. Counts and statistics based on these counts (factoring in DNA sequencing depth and library sizes) were used along with a cutoff to calculate a binary designation (enriched/not enriched) for each disynthon/condition pair. The conditions included: target only, target and competitive inhibitor, and no target (matrix only) control. Additionally, a binary indicator of promiscuity was calculated through an historic analysis of dozens of targets. First, a promiscuity ratio for each disynthon was calculated by taking the number of protein targets selected via immobilization on any Nickel IMAC resin where that disynthon was enriched and dividing by the total number of targets selected via immobilization on any Nickel IMAC resin where that disynthon had been screened to date. Then, a cutoff was applied and any disynthons with a higher ratio were considered promiscuous binders.  Altogether, this procedure resulted in the assignment of each disynthon to one of five classes: competitive hit, non-competitive hit, promiscuous binder, matrix binder, or non-hit. Competitive hits (the ``positive'' class for machine learning) included disynthons that (1) were enriched in the target condition, (2) were not enriched in the ``matrix only'' and ``target and competitor'' conditions, and (3) demonstrated a low promiscuity ratio.

\subsubsection*{Random forest models}
The training data were divided into training and test sets. The number of competitive binder training examples used for the RF models that were experimentally validated were \num{100000}, \num{87729}, and \num{100000} for sEH, ER$\alpha$, and c-KIT respectively. Test set size and composition varied, with sEH, ER$\alpha$, and c-KIT sets containing approximately \num{100000}, \num{10000}, \num{1000} positive examples, and \num{190000}, \num{125000}, \num{10000} negative examples respectively.  To address memory limitations during fitting, RF models were trained using 10 different random samples of competitive binder examples (positive examples were each included twice in the training set) in combination with four random samples of \num{500000} negative examples, resulting in a total of 40 different training sets. Fingerprint representations for all molecules were generated using the RDKit\cite{landrum2006rdkit} implementation of 1024-bit binary Morgan Circular Fingerprints with radius 2 (ECFP4)\cite{rogers2010extended}. Models were trained using the \texttt{RandomForestClassifier} class in scikit-learn\cite{pedregosa2011scikit}, with the following non-default hyperparameters: \texttt{n\_estimators=1000}, \texttt{min\_samples\_split=5}, \texttt{n\_jobs=6}, \texttt{max\_features='sqrt'}, \texttt{random\_state=42}. Performance was defined as the enrichment over random chance of positive examples (examples predicted at $\ge0.5$) in the test set. For each target, the top performing model was used to select predicted hits for experimental validation.

\subsubsection*{GCNN models}

\paragraph{Architecture.}
The GCNN was a ``weave'' graph-convolutional neural network, specifically the ``W2N2'' variant with input features and hyperparameters as specified by Kearnes \emph{et al.}~\cite{kearnes2016molecular} While the final linear layer in that work was used to make multi-task binary classification predictions, here the final linear layer was used to make predictions on the five mutually exclusive classes described above, trained with softmax cross entropy loss.

\paragraph{Cross-validation.}
A $k$-fold cross validation scheme which split the DEL data into train, tune, and test splits was used for the GCNN model. Each of the $k$ folds was specified as a grouping of one or more of the DNA-encoded libraries. The groupings of the libraries into folds were determined by plotting the first three Principal Components of the ECFP6 2048-bit binary vectors of a random sample of disynthons from each DNA-encoded library. After plotting, the libraries that clustered visually were grouped into the same fold, with ambiguities being resolved by grouping together libraries with similar combinatorial chemistry reactions. $k-2$ folds were then used for training each fold of the GCNN, with one tune fold reserved for training step selection, and one test fold reserved (but ultimately not used in this study). In the c-KIT and ER$\alpha$ models, 10\% of all the DEL selection data stratified by each of the 5 classes (randomly sampled by hash of molecule ID) were reserved as an ``ensemble holdout'' set (see Extended Data \figurename~\ref{fig:cross_validation}(b)). Due to third party use restrictions, for the ER$\alpha$ and c-KIT models a handful of productive DNA-encoded libraries were withheld from the GCNN training data, but they were used in fitting the Random Forest model. The number of competitive binder training examples used for the GCNN models that were experimentally validated were \num{355804}, \num{74741} and \num{50186} for the sEH, ER$\alpha$, and c-KIT targets respectively.

\paragraph{Oversampling during training.}
The vast majority of the training data is in the NON\_HIT class, and the cross-validation folds varied substantially in size. To improve training convergence time and stability of the GCNN, oversampling of the under-represented classes and cross-validation folds was used. The mechanism of oversampling was to constrain each stochastic gradient descent mini-batch to have equal numbers of disynthons from each class and cross-validation fold. Some fold/class combinations had fewer than 10 disynthons and were not used. Thus, mini-batch sizes varied slightly by cross-validation fold and protein target: the mini-batch size was the number closest to 100 that was evenly divisible by the number of fold/class combinations with at least 10 disynthons.

\paragraph{Step selection and ensembling.}
After training one model for each cross-validation fold, the model weights at the training step with the maximum ROC-AUC\cite{fawcett2006introduction} for the competitive hits class on the tuning set were selected. To generate model predictions on the Mcule and virtual library datasets for experimental validation, the median prediction for the compound across cross-validation fold models was used.

\paragraph{Performance.}
The average cross-validation ROC-AUC was $\sim0.8$. The ensembled model for c-KIT and ER$\alpha$ evaluated on the ``ensemble holdout'' reached a ROC-AUC of $\sim0.99$. See Extended Data Figure~\ref{fig:cross_validation} for details.

\subsubsection*{Compounds selected by similarity search}

To further determine contribution of machine learning on our ability to select potent molecules, a parallel experiment using Tanimoto similarity to positive training examples was conducted. Training structures were chosen from the pool of structures used in generation of GCNN models for ER$\alpha$, detailed as follows. Directed sphere exclusion\cite{gobbi2003dise} was used with Tanimoto similarity (ECFP6) cutoff of 0.35, ranked by the degree of enrichment in the target selection and the exemplar with highest enrichment from each of 994 clusters was chosen. The Mcule catalog was then searched for similars to the 994 training examples (molecules with >15 business day delivery time were excluded). Results were filtered to include compounds with ECFP6 Tanimoto scores of $\ge 0.55$. Directed sphere exclusion was again applied to the original list of Mcule similars using an ECFP6 Tanimoto cutoff of 0.35 and ranking by maximum similarity to the training examples. From each of the resulting 114 clusters, the exemplar with the highest similarity to any input molecule was chosen. 107 compounds were received and tested. This method produced no molecules with detectable activity.

\subsubsection*{Selection of diverse predicted compounds}

Selection of compounds for order or synthesis was made for each model from those with a prediction score over a specified cutoff (GCNN: 0.8, RF: 0.7 for Mcule and 0.5 for XVL) from either the Mcule catalog or from the XVL. 
 Removal of duplicated scaffolds (generated using the 'RDKit Find Murcko Scaffolds' Knime node) was performed on some predictions, retaining the more highly predicted structure. For GCNN Mcule selection, directed sphere exclusion clustering with ranking by model prediction score was applied using ECFP6 Tanimoto similarity with cutoffs determined empirically to reduce the number of molecules to hundreds or low thousands (GCNN Mcule sEH: 0.3, c-KIT: 0.5, ER$\alpha$: 0.45). For both RF and GCNN Mcule selection, hierarchical clustering was used as needed to further reduce to approximately 150 clusters. The most highly predicted compound was selected from each cluster. For Mcule orders, compounds weighing  $>$700 Dalton, less than a minimum MW ranging from 190-250 Daltons (varied by target and model), and/or those with too few heavy atoms ($\le$ 10) were removed.
 Molecules containing silicon were removed.
For all orders except sEH GCNN, Mcule molecules reporting delivery times of greater than 14 business days were excluded. 
To limit depletion of stocks, XVL compounds were filtered to limit the use of any single building block; the compound with the highest prediction score for any given building block was selected. To avoid synthesis problems, XVL compounds with reactants containing multiple reactive groups (e.g. two carboxylic acids) were removed.  For sEH XVL predictions, the top 150 remaining compounds were chosen and an additional 105 compounds were chosen by binning prediction scores into 21 bins (size 0.05, between 0.8 and 1.0) and choosing 5 randomly from each bin. The ``Match\_PAINS.vpy'' script provided with Dotmatics Vortex was applied for some compound purchase and synthesis requests. For both Mcule and XVL, an additional non-systematic visual filtering was performed by a chemist with or without the aid of substructure searches that was restricted to removal of molecules with the potential for instability or reactivity.

\subsubsection*{Molecular similarity comparisons}
Quantification of molecular structure similarity used Tanimoto similarity on extended-connectivity fingerprints\cite{rogers2010extended} with radius~3 (ECFP6). In this work we use a count-based representation (to better capture differences in molecular size with repeated substructures compared to binary fingerprints) and unhashed fingerprints (to avoid hash collisions).
ECFP6-counts vectors were generated with RDKit \cite{landrum2006rdkit} using the \texttt{GetMorganFingerprint()} method with \texttt{useCounts=True} argument.
Functional-Class Fingerprints (FCFP) are related to ECFP, but atoms are grouped into functional classes such as ``acidic'', ``basic'', ``aromatic'', etc before substructures are enumerated\cite{rogers2010extended}. Molecules which are similar structurally but have substitutions of similar atoms will look much more similar with FCFP than ECFP. FCFP6-counts (also with radius~3) were generated with \texttt{GetMorganFingerprint()} with \texttt{useCounts=True} and \texttt{useFeatures=True} arguments. Tanimoto similarity for two counts vectors (also commonly referred to as ``1 - Jaccard Distance'') is defined as the sum of the element-wise minimum of their counts divided by the sum of the element-wise maximum of their counts. A similarity value of 1.0 indicates identical structures (ignoring chirality), while 0.0 means that no substructures are shared. Nearest neighbors for hits in the training data were found using brute force exact search\cite{wu2019efficient} over the fingerprints.

\subsubsection*{Deep neural network architecture choice}
The experimentally validated results reported in this manuscript were derived from models trained on CPUs. GCNN models were trained to convergence on 100 CPU replicas for each fold, taking about a week for each model. Fully-connected deep neural networks (DNN) models trained on ECFP4\cite{rogers2010extended} bit vectors were considered for experimental validation but did not perform as well as GCNN in cross-validation. Extended Data \figurename~\ref{fig:cross_validation} compares cross-validation performance of GCNN and DNN models (with ReLu-activated layers of size 2000, 100), as quantified by ROC AUC\cite{fawcett2006introduction}. The cross-validation results in panel \textbf{(a)} of Extended Data \figurename~\ref{fig:cross_validation} come from models not used in this study's experimental results. They were trained on Tensor Processing Units\cite{jouppi2017datacenter}, on which the DNN and GCNN models converged in 2--3 hours, and the AUC reported is the mean AUC from 10 models trained from scratch with different random seeds. Each of the 10 models converged 8 independently randomly initialized sets of model weights and used the mean of the predictions from these 8 sets of weights as their overall prediction.

\subsubsection*{ChEMBL searches for published inhibitors}

For sEH, a search for ``epoxide hydrolase'' was conducted through the ChEMBL\cite{gaulton2016chembl} website at \url{https://www.ebi.ac.uk/chembl/}. Targets were narrowed by organism to \emph{homo sapiens}, and target entries for other proteins were removed. Bioactivity results were retrieved for the relevant target entry. Results were limited to K$_i$, K$_d$ and IC$_{50}$ values (\emph{i.e.} percent inhibition values were removed). All values qualified with `>' or `$\ge$' were removed, as were compounds reported with K$_i$, K$_d$ and IC$_{50}$ >10~{\textmu}M. All remaining (1607 compounds) were used for similarity comparison. Target specific searches were conducted for ER$\alpha$ ('Estrogen Receptor') and c-KIT ('KIT'); identification of published actives followed this same procedure producing 2272 and 1288 compounds respectively.

\subsubsection*{Reproducibility of training data}
Two DEL selections were performed on sEH months apart. Disynthon aggregation and labeling as described above resulted in training labels (as determined by thresholded enrichment values) that cross-predicted each other almost perfectly. We quantified this cross-prediction performance by calculating the Area Under the Curve (AUC) of the Receiver Operator Characteristic (ROC) curve\cite{fawcett2006introduction}. Using the first DEL selection's positive-class enrichment values as a ranking function to predict the positive-class binary training label of the second DEL selection achieved a ROC-AUC equal to 0.97, and predicting the first DEL selection's training label from the second DEL selection's enrichment values achieved 0.99 (See Extended Data \figurename~\ref{fig:two_selections_roc_curve}).

\subsection*{Experimental}

\subsubsection*{On-demand synthesis of virtual library compounds}
A virtual library (XVL) comprising 83.2~M compounds was enumerated as the product of amide formation of all compatible building blocks available in the X-Chem in-house inventory.  Small libraries of compounds chosen via machine learning prediction and filtering were synthesized in parallel on a micromole scale (about 1~{\textmu}mol).  The synthesis was performed in 96 well plates using a conventional synthesis protocol with DMT-MM as the coupling agent.  The crude reaction mixtures were filtered through filter plates fitted with an alumina plug.  The semi-purified reaction mixtures were analyzed using LC-MS to evaluate the reaction efficiency.  The eluents were collected in 96 well receiving plates and diluted to 1~mM solution in DMSO that was used directly for the primary biochemical assay. A small number of XVL compounds (4) identified by both GCNN and RF models for ER$\alpha$ were synthesized and tested independently for each model and are reported separately in the figures and supplementary data.

\subsubsection*{Affinity‐mediated selection}

All affinity-mediated selections included between 31 and  42 DEL libraries synthesized as described in Cuozzo \textit{et al.}\cite{Cuozzo2017Disc}. For each target, purified protein (sEH: 1 {\textmu}M, c-KIT (wild type): 3 {\textmu}M, ER$\alpha$ (wild type): 8 {\textmu}M) each  containing a His6 tag were incubated
in solution with DNA-encoded library (40 {\textmu}M) for 1 hour in a volume of 60 {\textmu}L in 1x selection buffer. 1x selection buffer
consisted of HEPES (20 mM), potassium acetate (134 mM), sodium acetate (8 mM), sodium chloride (4 M), magnesium acetate
(0.8 mM), sheared salmon sperm DNA (1 mg/mL, Invitrogen AM9680), Imidazole (5 mM), and TCEP (1 mM) at pH 7.2.
1x selection buffer for sEH additionally included Pluronic F-127 (0.1\%) and 1x selection buffer for ER$\alpha$ and
c-KIT additionally included Tween 20 (0.02\%). For each target, an additional selection condition containing both target and
40–100 {\textmu}M of a competitive inhibitor of the target was run in parallel. The competitive inhibitor was pre-incubated with the target in 1x selection
buffer for 0.5 hour prior to addition of the DNA-encoded library. For each target, an additional selection condition containing
no target was run in parallel. For each selection condition (no target, target or target with competitive inhibitor), a separate ME200 tip
(Phynexus) containing 5 {\textmu}L of nickel affinity matrix was pre-washed 3 times in 200 {\textmu}L of appropriate, fresh 1x selection buffer.
The affinity matrix used for sEH and c-KIT was HIS-Select HF Nickel Affinity Gel (Sigma H0537) and the affinity matrix
used for ER$\alpha$ was cOmplete\textsuperscript{TM} His-Tag Purification Resin (Sigma 5893682001). Each selection was separately
captured with 20 passages over the appropriate ME200 tip for a total of 0.5 hour. The bound protein/library captured on the
ME200 tip was washed 8 times with 200 {\textmu}L of appropriate, fresh 1x selection buffer. Bound library members were eluted
by incubating the ME200 tip with 60 {\textmu}L of 1x fresh, selection buffer at 85°C for 5 min. The solution from the heat elution
was incubated with 20 passages over a fresh, pre-washed ME200 tip containing 5 {\textmu}L of nickel affinity matrix to remove any
eluted protein. This selection process was run a second time using the eluate of the first selection in place of the input DNA-encoded library and using no target, fresh target or fresh target with competitive inhibitor as appropriate. The eluate of the second round of selection
was PCR amplified in a volume of 200 {\textmu}L with 5’ and 3’ primers (0.5 {\textmu}M each) and 1x Platinum PCR Supermix (Invitrogen
11306-016) with 15–25 cycles of [denaturation 94°C for 30 sec, annealing 55°C for 30 sec, and extension 72°C for 120 sec] until the double-stranded amplification products were clearly visible on an ethidium-stained 4\% agarose gel.
These primers include Illumina READ1 or READ2 sequences as required for sequencing on an Illumina HiSeq
2500. PCR-amplified selection output was then sequenced on an Illumina HiSeq 2500. Sequence read numbers (in millions) of the selections ([target, no target control, target + competitive inhibitor]) were [93, 95, 90] for sEH, [41, 18, 39] for c-KIT, and [56, 31, 65] for ER$\alpha$. Sequence data were parsed, error-containing sequences were disregarded, amplification duplicates were removed and building block and chemical scheme encodings were decoded and reported along with associated calculated statistical parameters.

\subsubsection*{Biochemical assays}

\paragraph{sEH assay.}

The IC$_{50}$ values for soluble epoxide hydrolase compounds were determined using the biochemical activity assay described by Litovchick \emph{et al.}~\cite{litovchick2015encoded}

\paragraph{c-KIT wild type assay.}
The IC$_{50}$ values for c-KIT were determined using an ADP-Glo assay.  Recombinant kinase domain was diluted in assay buffer, 20~mM HEPES pH~7.5, 10~mM Mg acetate, 100~mM Na acetate, 1~mM DTT, 0.1\% Pluronic F127, such that the final assay concentration was 30~nM.  Serially diluted test compounds were then added to the assay plate. Both ATP and peptide substrate were then added to a final concentration of 100~{\textmu}M each. The reaction was incubated for 1 hour at room temperature and then terminated by the addition of ADP-Glo reagent and kinase detection reagents (Promega).  The final reaction volume was 12~{\textmu}L. A luminescence plate reader was used to measure the signal generated by the ADP-Glo reagents and the data points were plotted against compound concentrations.

\paragraph{ER$\alpha$ wild type assay.}
Two assays were used in the course of this work reflecting availability of two different reagents. Consistency of results between the two assays was validated with a reference compound. 

Inhibition values for ER$\alpha$ compounds were determined using a homogeneous time-resolved fluorescence energy transfer assay (HTRF). Recombinant GST-tagged ER$\alpha$ (Thermo Fisher Scientific) was diluted into nuclear receptor assay buffer (Thermo Fisher Scientific) containing a terbium-labeled anti-GST antibody (Thermo Fisher Scientific). Serial dilutions of test compounds dissolved in DMSO or DMSO-only controls were dispensed into the assay plate in a volume of 120~nL and then 6~{\textmu}L of GST-tagged ER$\alpha$/terbium anti-GST antibody was added to the wells and incubated for 15~minutes at room temperature. The final assay concentrations of GST-tagged ER$\alpha$/antibody were 2.1~nM and 2~nM respectively. A volume of 6~{\textmu}L fluorescent ligand was then added to each well to a final concentration of 3~nM and the plates were further incubated at room temperature for 4~hours to allow binding to reach equilibrium. HTRF signal was measured using an excitation wavelength of 337~nm and emission wavelengths of 490~nm/520~nm on a fluorescent plate reader. The 520~nm emission signal was normalized using the 490~nm signal and plotted against compound concentrations.

For assaying compounds chosen by the similarity search, we used a fluorescence polarization based protocol using recombinant His-tagged ER$\alpha$ (in-house generated). The final assay concentrations of His-tagged ER$\alpha$  and fluorescent ligand were 5~nM and 3~nM, respectively, in a total reaction volume of 12~{\textmu}L. Compounds were pre-incubated with receptor for 15~minutes at room temperature prior to addition of the fluorescent ligand. After further incubation for one hour, the fluorescence polarization signal was measured using an excitation wavelength of 485~nm and emission wavelength of 535~nm.

\subsection*{Assay cascade and reported potency values}
In the first round of experiments for each target, single-point inhibition assays were run, and those ligands meeting the thresholds listed in Extended Data Table~\ref{table:cutoffs} were re-tested with at least two 10-point dose--response curves. IC$_{50}$ values were calculated by fitting the data points to a sigmoidal curve using a four-parameter logistic model. To best utilize available budget for dose--response curves in this study, these thresholds were decided after the single-point assays were run, solely based on the number of molecules that would consequently receive dose--response testing. When reporting hit potencies and hit rates in figures and text of this work, we aggregated data from both single-point inhibition assays and full dose--response curves. All potencies reported as under 10~{\textmu}M are the geometric mean of at least two validated (10-point curve) IC$_{50}$ values. Dose--response curves were validated and IC$_{50}$ values excluded where the Hill slope of logistic fit < 0.5 or > 3.0 or R\textsuperscript{2} < 0.8 (when inhibition >50\% at max concentration) or R\textsuperscript{2} < 0.6 (when inhibition $\leq$50\% at max concentration). Hits reported as 30~{\textmu}M potency come from one of the following three categories: \textbf{1)} geometric mean of least two (10-point curve) IC$_{50}$ values was less than 30~{\textmu}M \textbf{2)} only one of the tested dose--response curves resulted in a valid IC$_{50}$ (ranging from 13~nM to 28.43~{\textmu}M)) or \textbf{3)} single-point inhibition assays (at 10~{\textmu}M or 30~{\textmu}M) showed > 50\% inhibition but the compound was not re-tested with full dose--response curves due to resource constraints.

\section*{Data availability}
Chemical structures and experimentally determined potency values for tested compounds are available in a supplementary CSV; 89 of 1992 (4.5\%) have been structure-anonymized due to similarity to molecular intellectual property related to either partnered or internal drug development programs.

\section*{Code availability}
scikit-learn RandomForestClassifier was used to generate the RF predictions. A custom tensorflow implementation of the Kearnes et. al. ``W2N2'' graph convolutional neural network\cite{kearnes2016molecular} was used to generate the GCNN predictions.

\bibliography{main}

\section*{Acknowledgements}

We acknowledge: AstraZeneca for providing reagents used in screening of both ER$\alpha$ and c-Kit; Zan Armstrong for help with visual design of figures; the
X-Chem Library Synthesis and Design teams for the DEL libraries; the X-Chem Scientific Computing team for analytical tools and database capabilities; the X-Chem Lead Discovery team for input and contributions to the X-Chem DEL tagging strategy, target screening and DEL selection analysis. Rick Wagner, Terry Loding, Allison Olsziewski, Anna Kohlmann, Jeremy Disch and Belinda Slakman for valuable input and support during this study and the writing of this manuscript.

\section*{Author contributions statement}

P.R. and E.A.S. conceived and directed the study.
K.M., S.K., E.A.S., L.X., C.J.M., Y.Z. developed the cross-validation scheme for GCNN and/or the disynthon classification scheme used in both models.
K.M. and S.K. trained the GCNN models.
D.M. and L.X. trained the RF models.
E.A.S., L.X. and C.D.H. applied pre-arranged structural filtering to model predictions.
X.T. performed compound synthesis and characterization of virtual library compounds.
D.G., S.B. and B.C. performed activity assays. 
M.-A.G. performed statistical calculations on DEL data.
A.D.K. identified suitable protein targets and selection output datasets.
J.P.G. performed DEL affinity-mediated selections.
E.A.S. and M.A.C. designed the similarity search baseline experiment.
K.M., S.K., E.A.S., C.J.M., C.H. performed analysis of activity assay results.
K.M., P.R., S.K., E.A.S., C.H., J.W.C., D.G., J.P.G., Y.Z., A.D.K. and C.J.M. wrote the paper.


\section*{Additional information}

\subsection*{Competing interests}
All authors are current or former employees of X-Chem, Inc. or Google LLC as noted in their author affiliations. X-Chem is a biotechnology company that operates DNA-encoded library technology as part of its business. X-Chem has filed a PCT application covering the use of DEL data with machine learning for hit identification (PCT/US2018/028050, inventors E.A.S., L.X., D.M., C.J.M.). Google is a technology company that sells machine learning services as part of its business. Portions of this work are covered by issued US Patent No. 10,366,324 ("Neural Network for Processing Graph Data", P.R. is an inventor) and a pending unpublished US patent application, both filed by Google. ZebiAI is a biotechnology company that applies machine learning to DEL selection data as part of its business. Cognitive Dataworks is a commercial consulting and software company.

\section*{Figures}

\begin{table}[!htb]
\centering
\rowcolors{2}{lightgray}{}
\begin{tabular}{c c S c S}
\toprule
{Target} & {Confirmed Hit} & {\makecell{IC$_{50}$ \\ (nM)}} & {\makecell{Nearest \\ ChEMBL Hit}} & {Similarity} \\
\midrule
sEH & \includegraphics[scale=0.6]{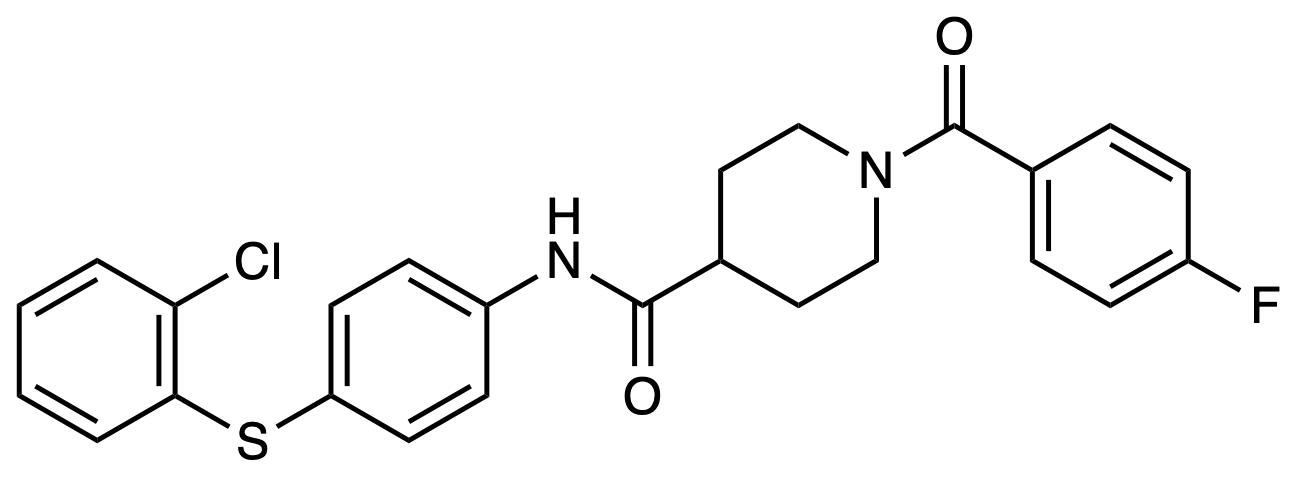} & 1 & \includegraphics[scale=0.6]{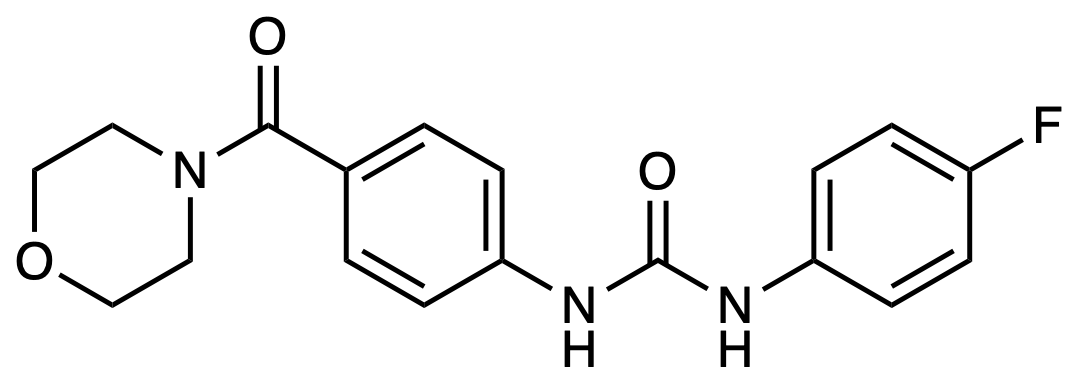} & {0.39 (0.4)} \\
sEH & \includegraphics[scale=0.6]{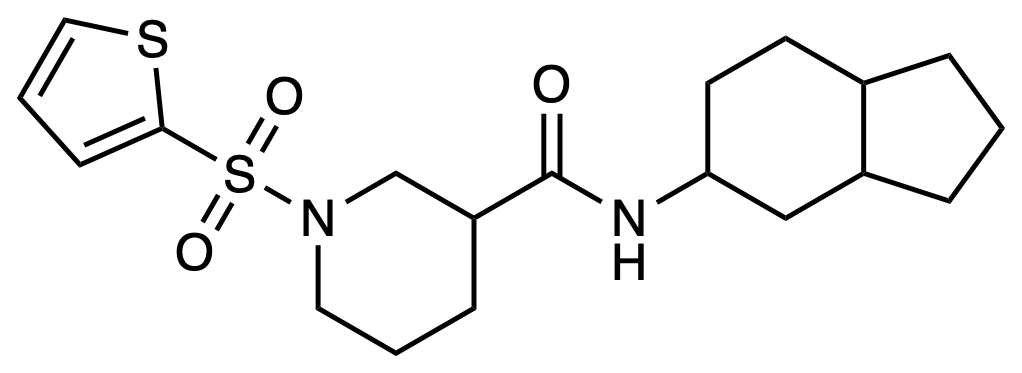} & 2 & \includegraphics[scale=0.6]{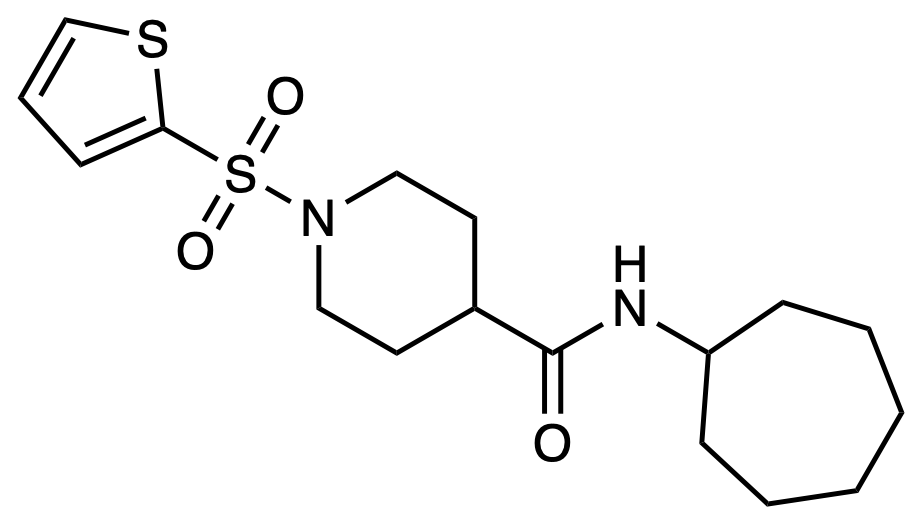} & {0.44 (0.65)} \\
ER$\alpha$ & \includegraphics[scale=0.6]{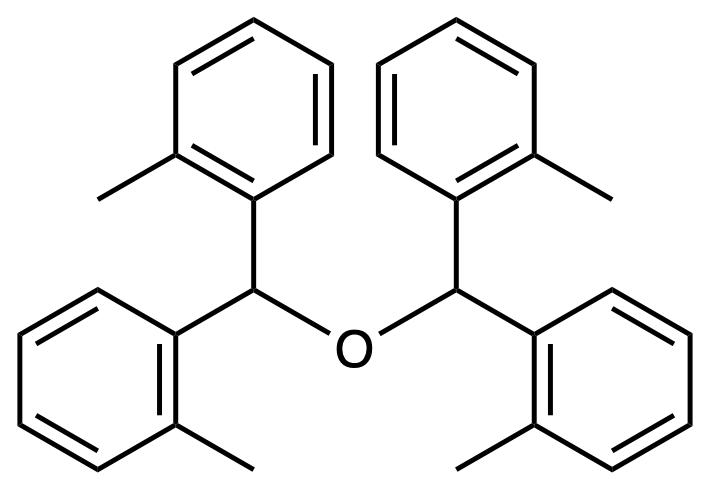} & 8 & \includegraphics[scale=0.6]{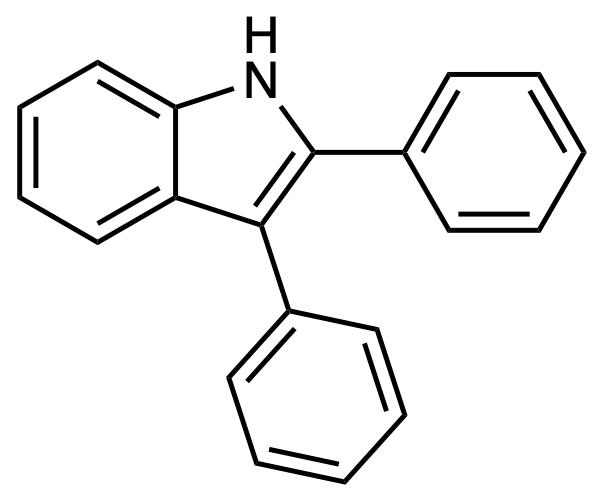} & {0.26 (0.28)} \\
ER$\alpha$ & \includegraphics[scale=0.6]{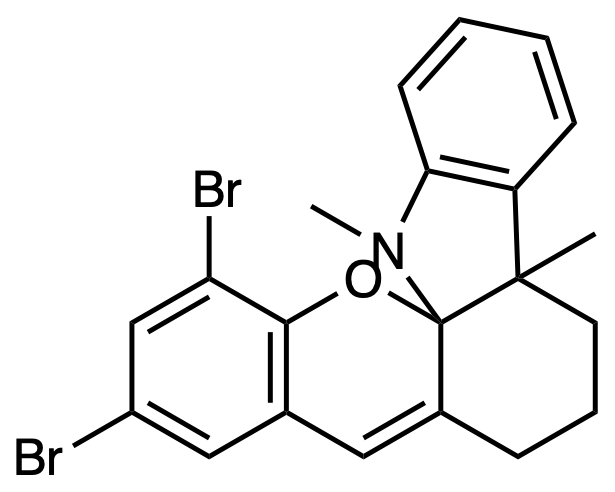} & 20 & \includegraphics[scale=0.6]{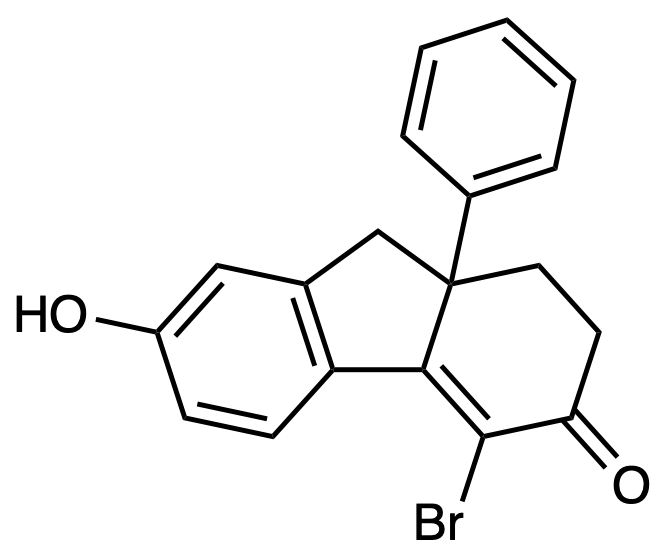} & {0.2 (0.27)} \\
c-KIT & \includegraphics[scale=0.6]{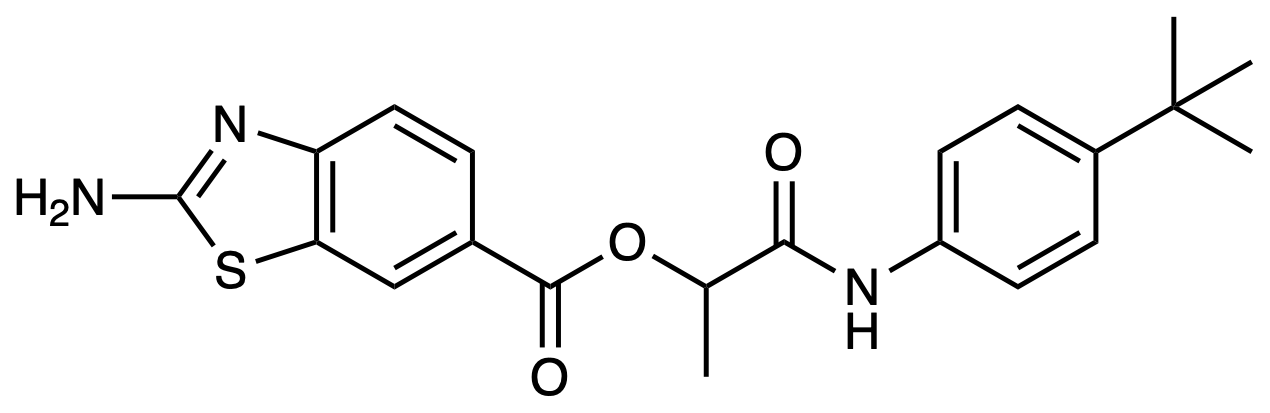} & 9 & \includegraphics[scale=0.6]{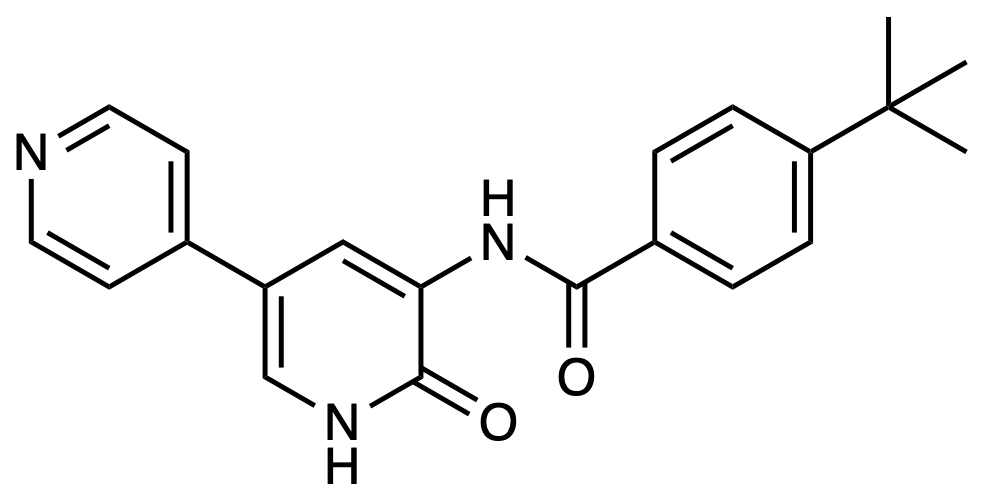} & {0.3 (0.33)} \\
c-KIT & \includegraphics[scale=0.6]{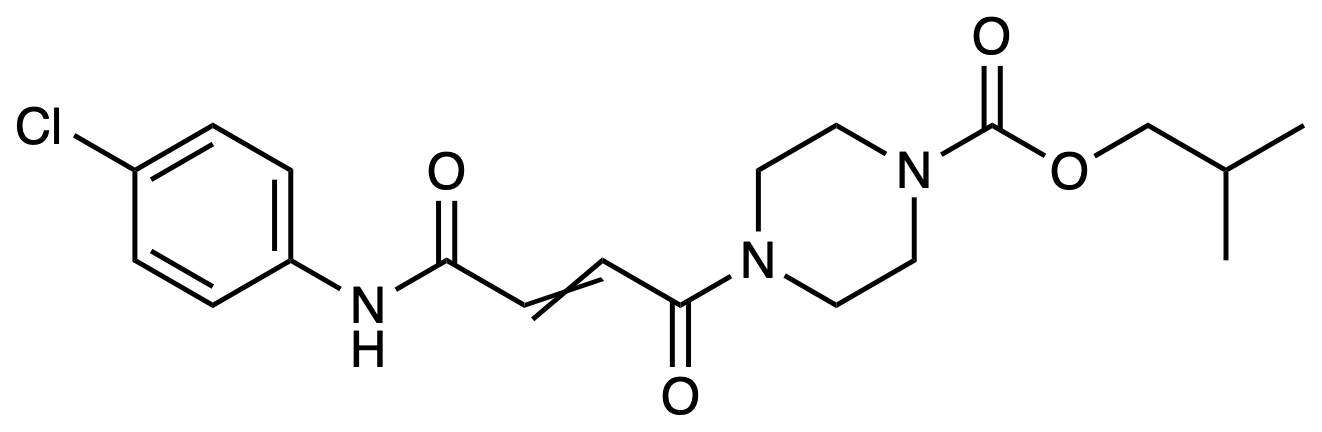} & 21 & \includegraphics[scale=0.6]{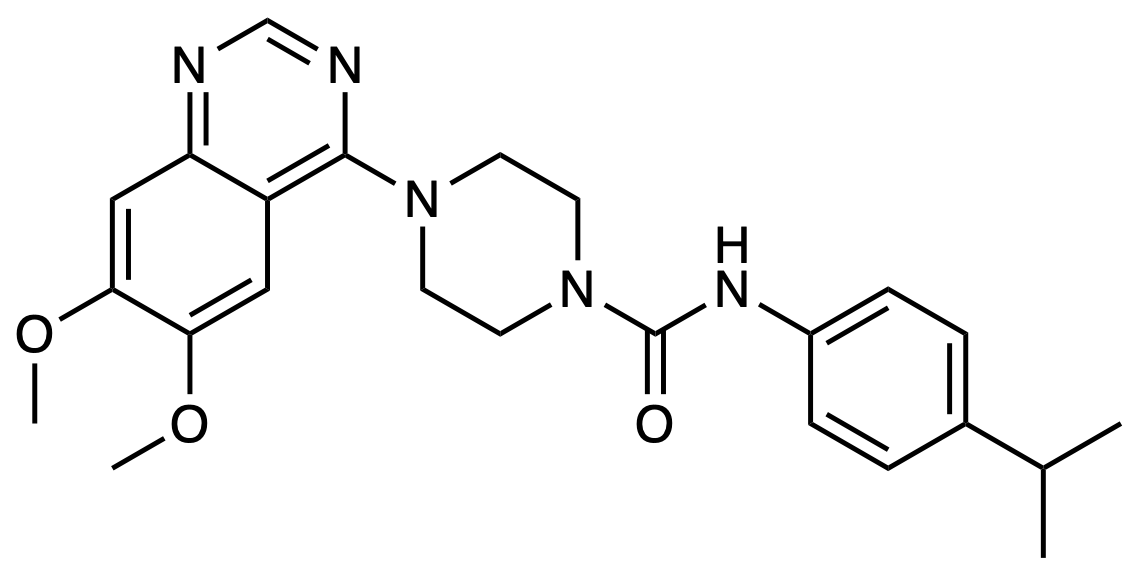} & {0.23 (0.25)} \\
\bottomrule
\end{tabular}
\caption{Examples of potent hits for each target. For each hit compound, we show the closest previously known ChEMBL hit as measured by Tanimoto on ECFP6-counts fingerprints. Similarity values are given as ECFP6-counts (FCFP6-counts). A redacted set of hits and nearest neighbors for all targets is given in the Supplementary Information.}
\label{table:potent_hits}
\end{table}

\clearpage

\renewcommand{\figurename}{Extended Data Figure}
\setcounter{figure}{0}
\renewcommand{\tablename}{Extended Data Table}
\setcounter{table}{0}

\begin{figure}[t]
\centering
\includegraphics[width=\linewidth]{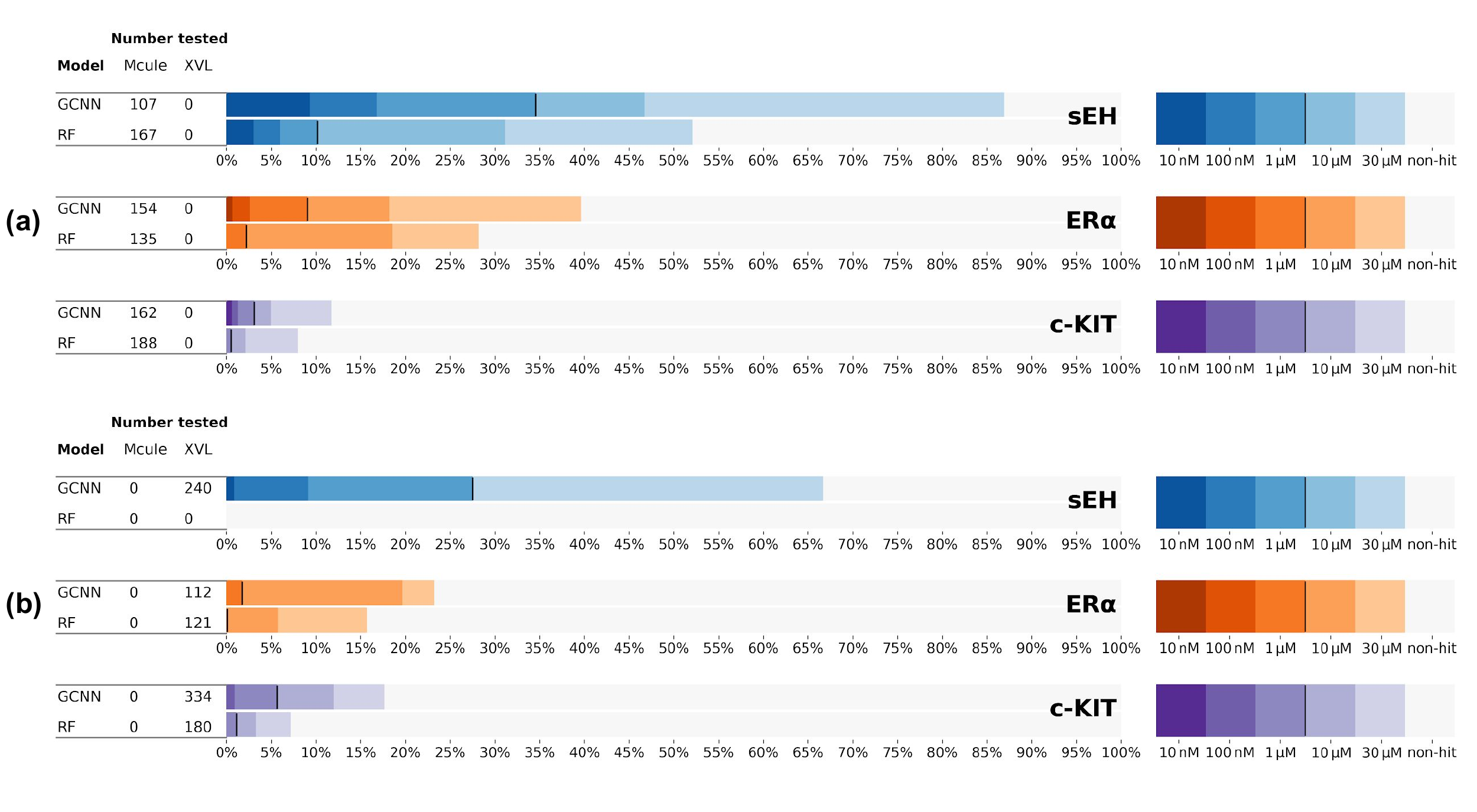}
\caption{Hit rates and potencies broken out by Mcule (\textbf{a}) and XVL (\textbf{b}) compounds.}
\label{fig:hit_rates_mcule_and_vl1}
\end{figure}

\begin{figure}[t]
\centering
\includegraphics[width=\linewidth]{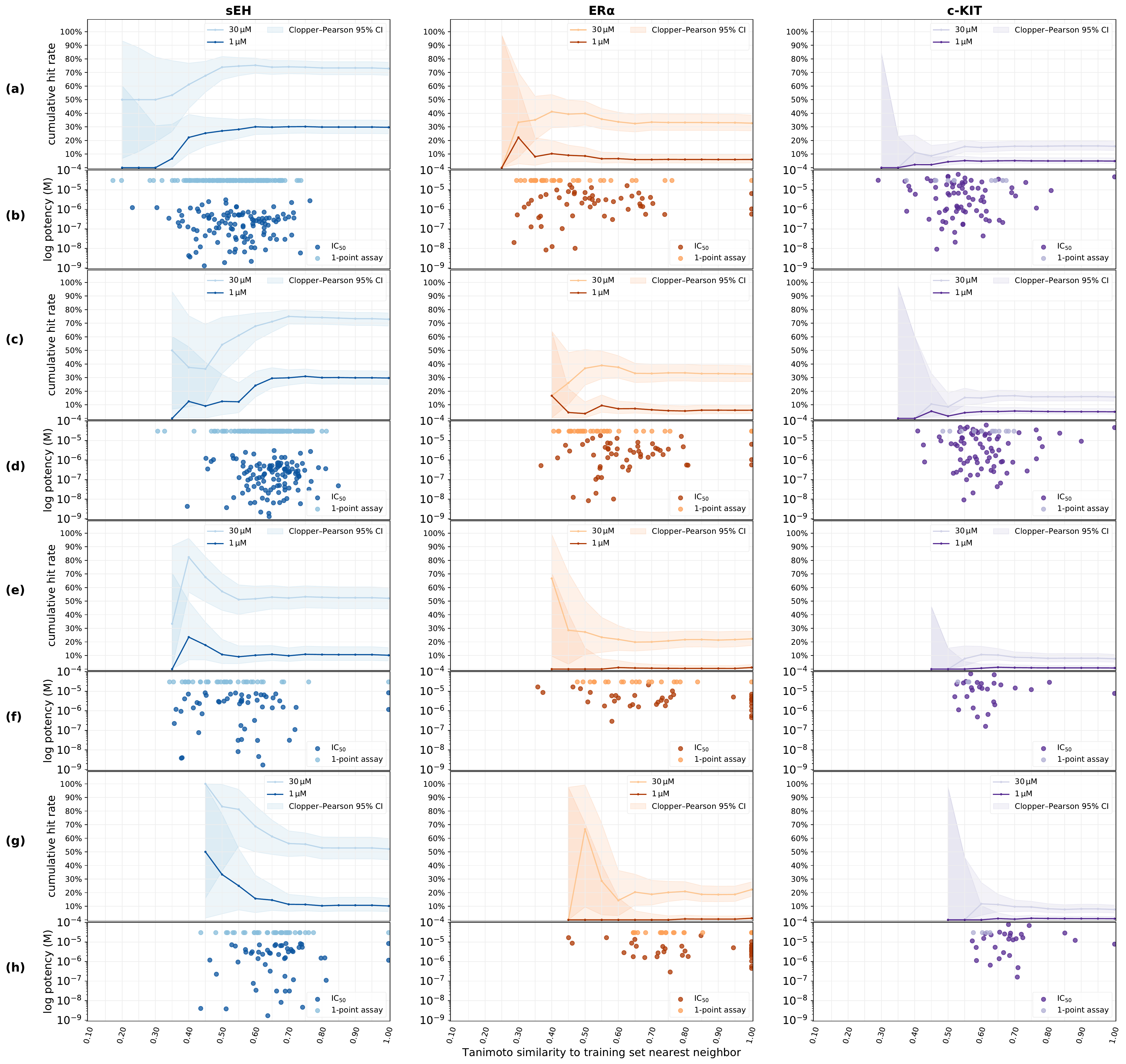}
\caption{Cumulative hit rates of model-predicted compounds \textbf{(a), (c), (e), (g)}, along with scatter plots of hits \textbf{(b), (d), (f), (h))} on a shared x-axis of Tanimoto similarity of compounds to the training DELs. GCNN-predicted compounds in \textbf{(a), (b)} use ECFP6-counts fingerprints for x-axis similarity.
GCNN-predicted compounds in \textbf{(c), (d)} use FCFP6-counts fingerprints for x-axis similarity. RF-predicted compounds in \textbf{(e), (f)} use ECFP6-counts fingerprints for x-axis similarity. RF-predicted compounds in \textbf{(g), (h)} use FCFP6-counts fingerprints for x-axis similarity. The cumulative hit rate plots show the hit rates for compounds with $\leq$ a given (x-axis) similarity to the training set. Error bands are Clopper--Pearson intervals\cite{clopper1934use} at 95\% confidence.}
\label{fig:hit_rate_by_sim_extended}
\end{figure}

\begin{figure}[t]
\centering
\includegraphics[width=\linewidth]{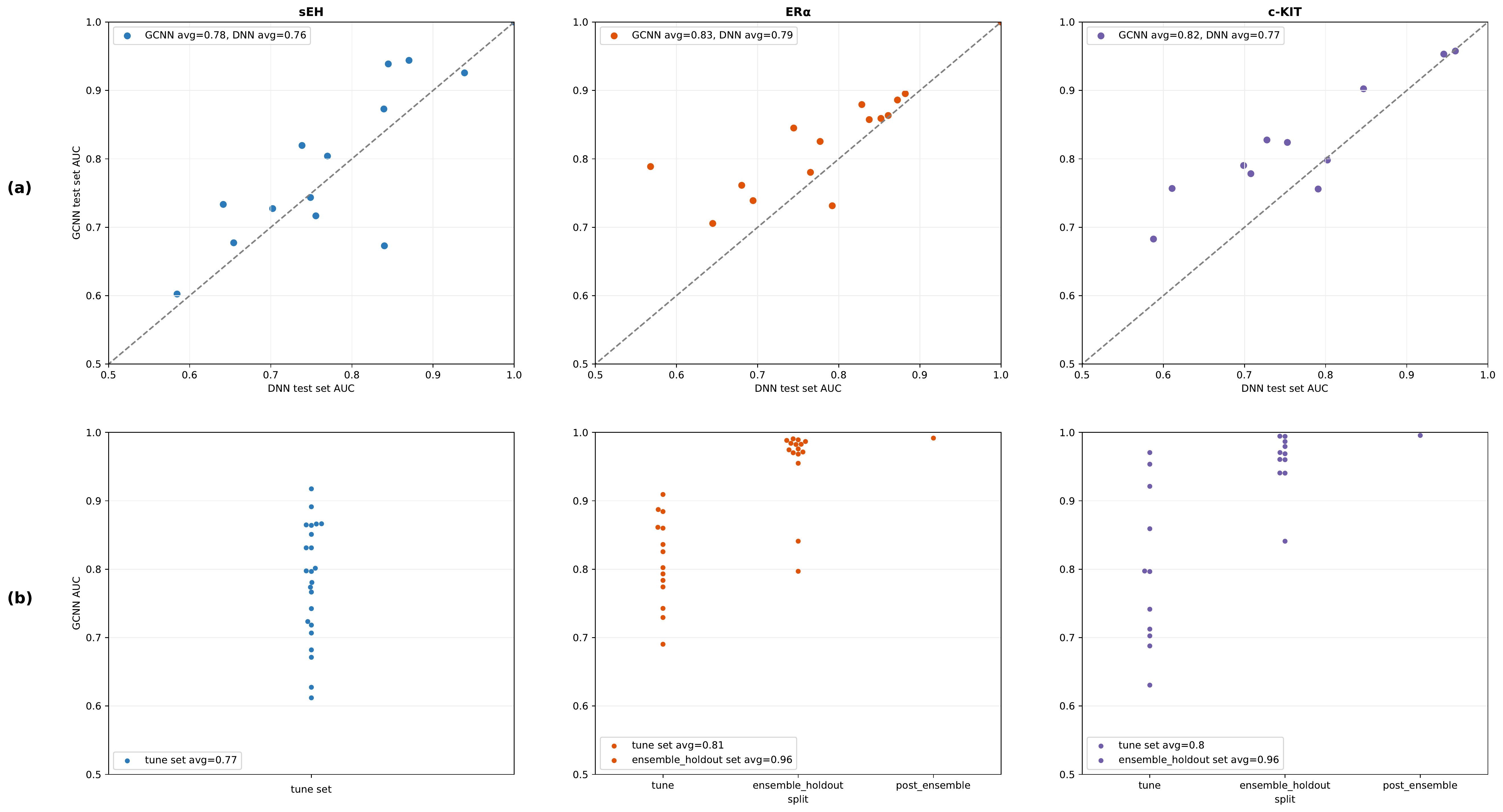}
\caption{Comparison of cross validation model performance, between \textbf{(a)} a GCNN model and a DNN model (see Methods), and \textbf{(b)} on holdout sets evaluated on the models used to make experimentally validated predictions.}
\label{fig:cross_validation}
\end{figure}

\begin{figure}[t]
\centering
\includegraphics[width=\linewidth]{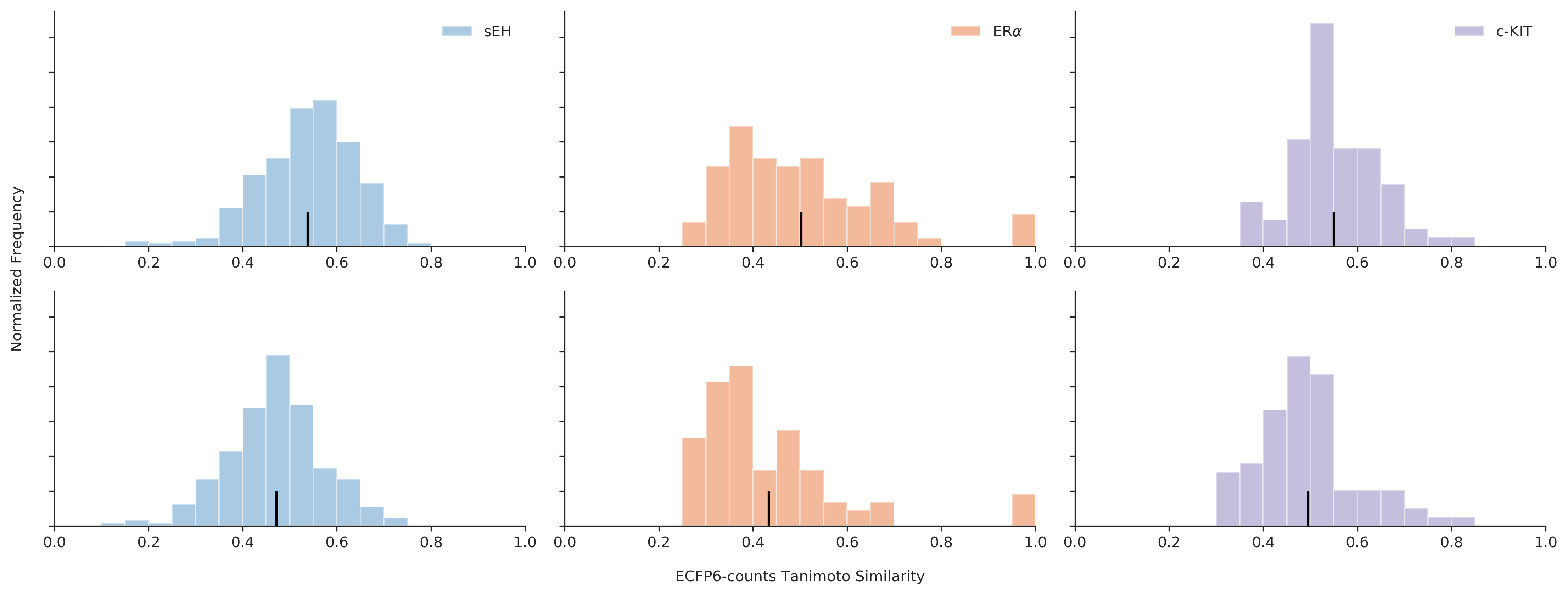}
\caption{Similarity between confirmed hits and nearest training examples from GCNN predictions. \textbf{(Top row)} Distributions of ECFP6-counts Tanimoto similarity between confirmed hits and the most similar compound in the training set. \textbf{(Bottom row)} Distributions of ECFP6-counts Tanimoto similarity between confirmed hits and the most similar positive training example. In all plots, the distribution mean is indicated with a vertical black line.}
\label{fig:hits_neighbors}
\end{figure}

\begin{figure}[t]
\centering
\includegraphics[width=\linewidth]{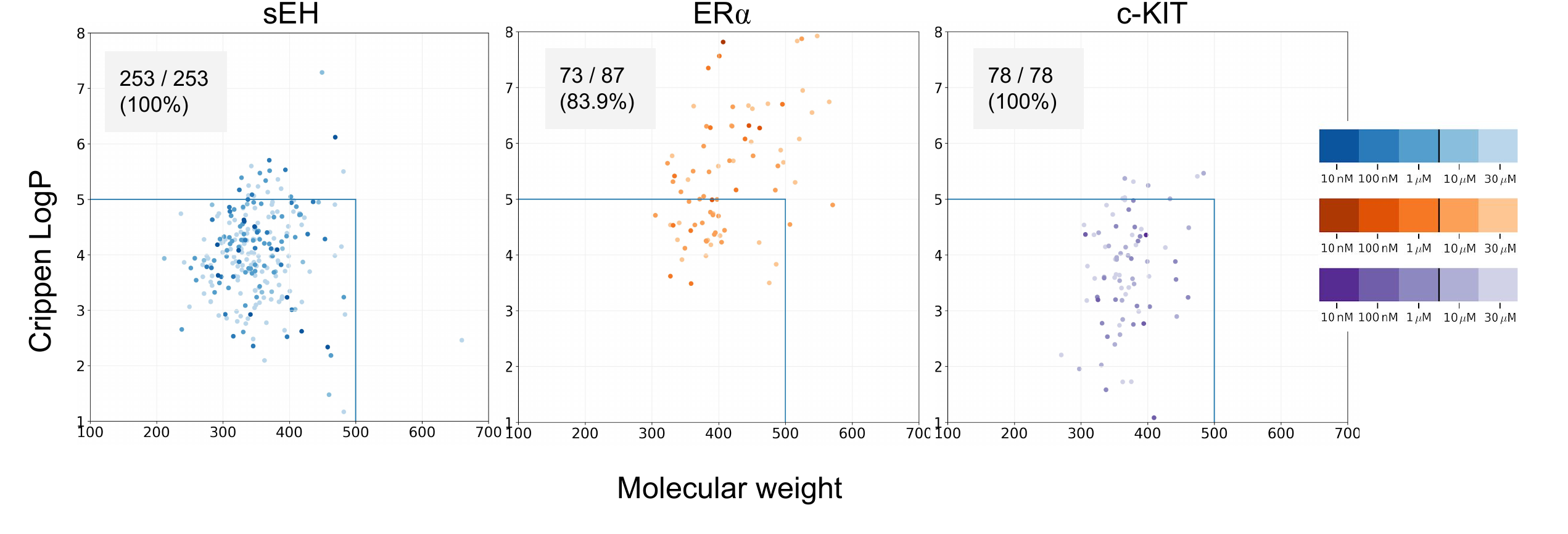}
\caption{Scatter plot of molecular weight and calculated Crippen LogP for confirmed active ligands predicted by GCNN. Blue lines indicate Lipinski "Rule of 5" thresholds. Inset in gray boxes are the portion of hits at 30~{\textmu}M that have 1 or fewer "Rule of 5" violations.}
\label{fig:lipinski}
\end{figure}

\begin{figure}[!htb]
\centering
\includegraphics[width=\linewidth]{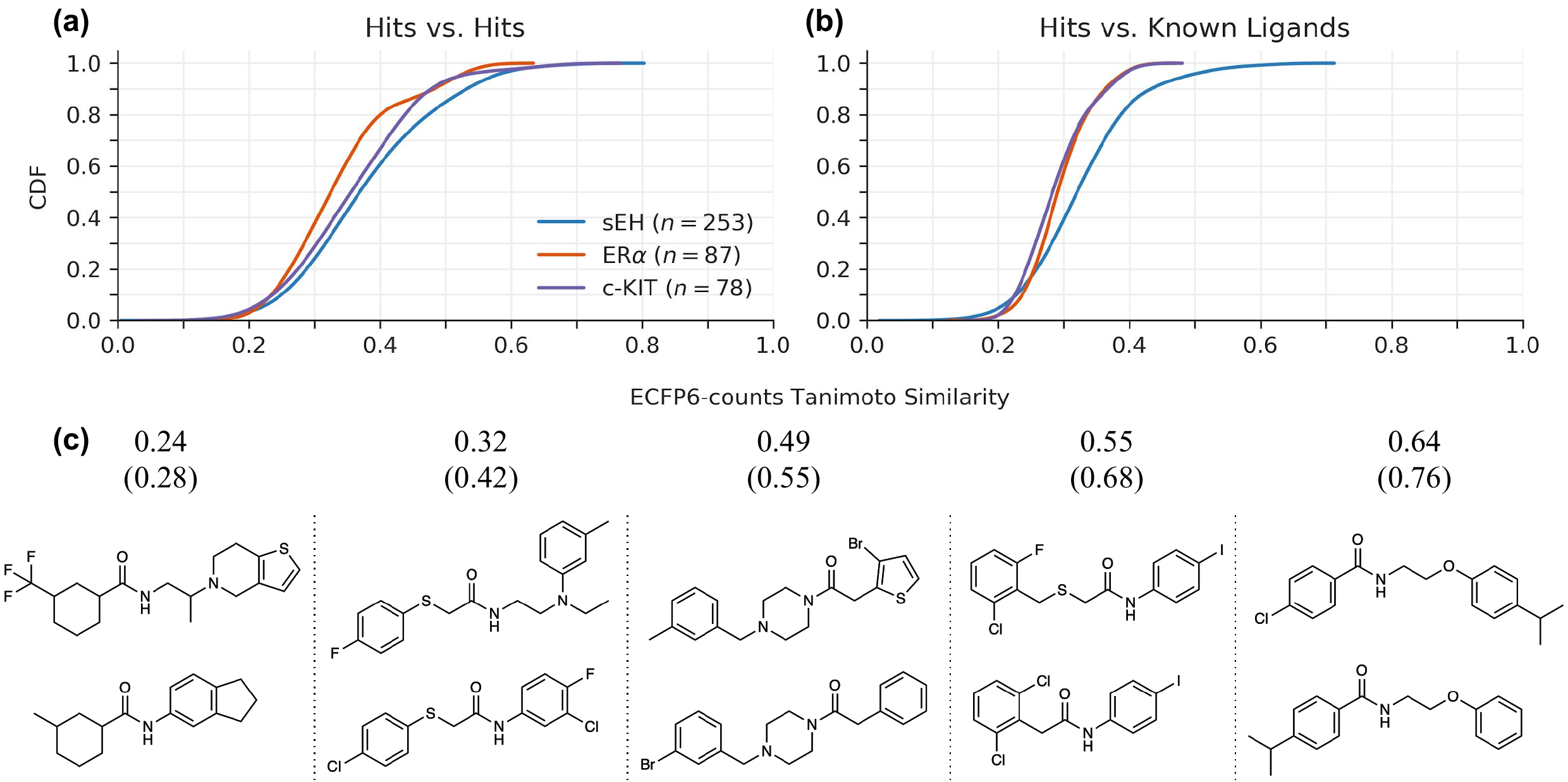}
\caption{Similarity between confirmed GCNN hits. \textbf{(a, b)} Cumulative distribution functions (CDFs) of maximum ECFP6-counts Tanimoto similarity between compounds for each target. For each compound, the maximum similarity to (a) other hits or (b) known ligands for the same target is reported. The number of known ligands for each target is as follows: 1607 (sEH), 2272 (ER$\alpha$), 1288 (c-KIT). A redacted set of hit structures and the full set of known ligands for each target are available as Supplementary Information. \textbf{(c)} Examples of hit--hit pairs for sEH are shown to illustrate similarity at a variety of Tanimoto levels; similarity values are given as ECFP6-counts (FCFP6-counts).}
\label{fig:hit_vs_hit}
\end{figure}

\begin{table}[htb]
\centering
\rowcolors{2}{lightgray}{}
\begin{tabular}{c c S c S c S}
\toprule
{Target} & {Confirmed Hit} & {\makecell{IC$_{50}$ \\ (nM)}} & {\makecell{Nearest Training \\ Positive (NTP)}} & {\makecell{Similarity \\ to NTP}} & {\makecell{Nearest Training \\ Neighbor (NTN)}} & {\makecell{Similarity \\ to NTN}} \\
\midrule
sEH & \includegraphics[scale=0.6]{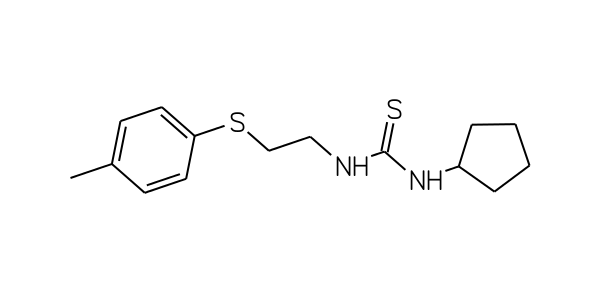} & 853 & \includegraphics[scale=0.6]{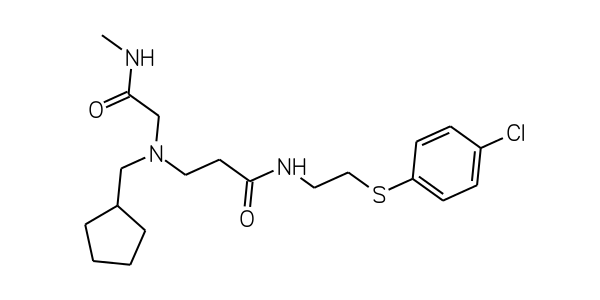} & {0.35 (0.28)} & \includegraphics[scale=0.6]{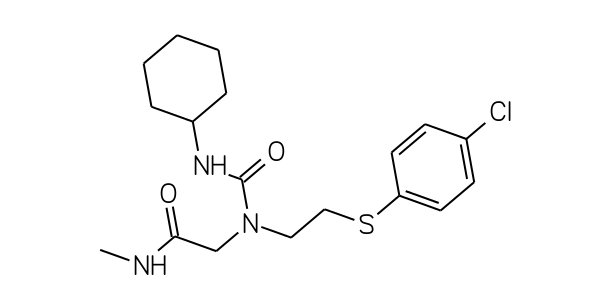} & {0.37 (0.3)} \\
sEH & \includegraphics[scale=0.6]{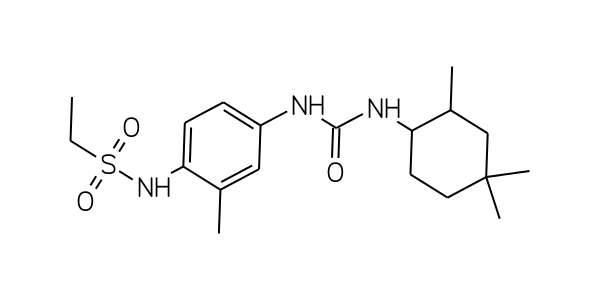} & 4 & \includegraphics[scale=0.6]{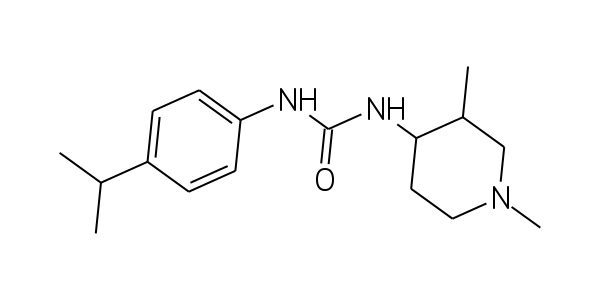} & {0.3 (0.41)} & \includegraphics[scale=0.6]{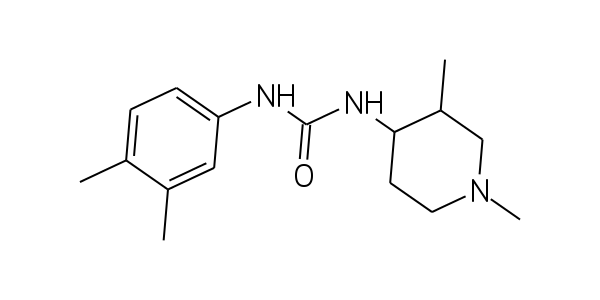} & {0.4 (0.4)} \\
sEH & \includegraphics[scale=0.6]{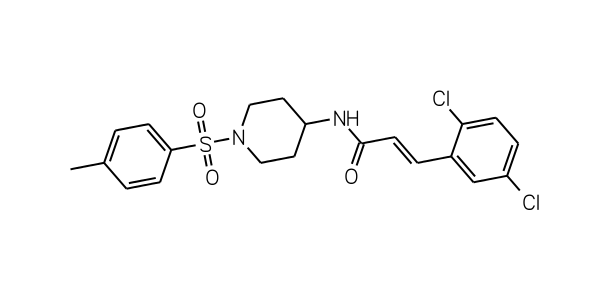} & 95 & \includegraphics[scale=0.6]{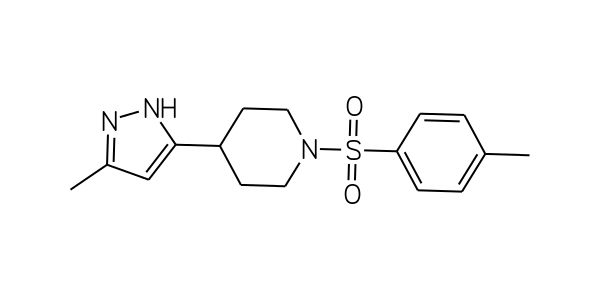} & {0.41 (0.43)} & \includegraphics[scale=0.6]{rdkit_structures/XLIB0040___GL_043420_GL_037990.png} & {0.41 (0.43)} \\
sEH & \includegraphics[scale=0.6]{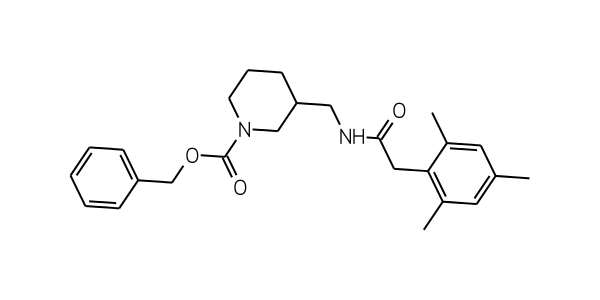} & 113 & \includegraphics[scale=0.6]{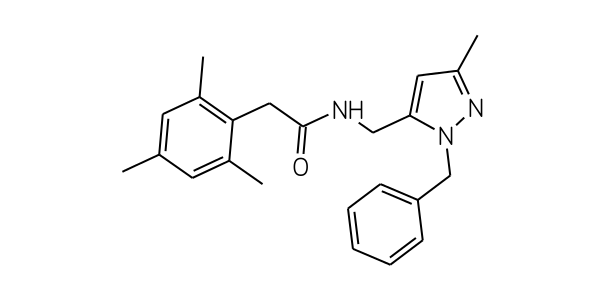} & {0.44 (0.44)} & \includegraphics[scale=0.6]{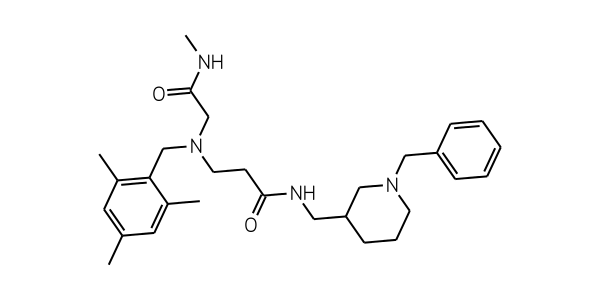} & {0.49 (0.44)} \\
ER$\alpha$ & \includegraphics[scale=0.6]{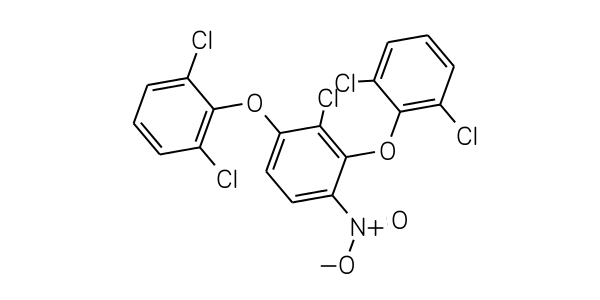} & 522 & \includegraphics[scale=0.6]{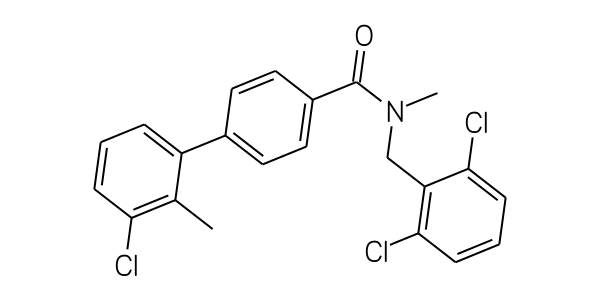} & {0.27 (0.29)} & \includegraphics[scale=0.6]{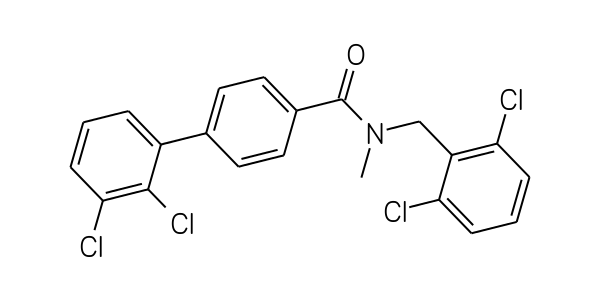} & {0.3 (0.31)} \\
ER$\alpha$ & \includegraphics[scale=0.6]{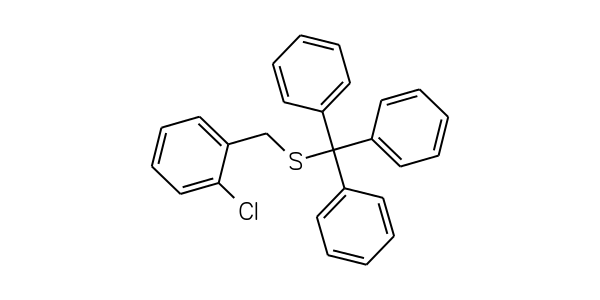} & 133 & \includegraphics[scale=0.6]{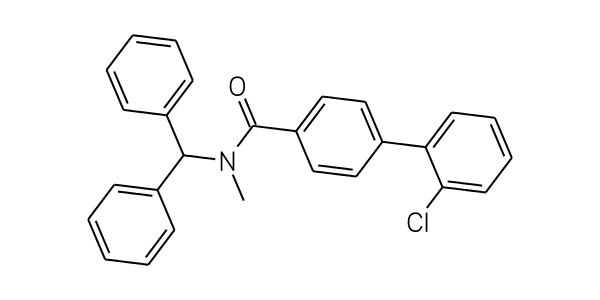} & {0.35 (0.51)} & \includegraphics[scale=0.6]{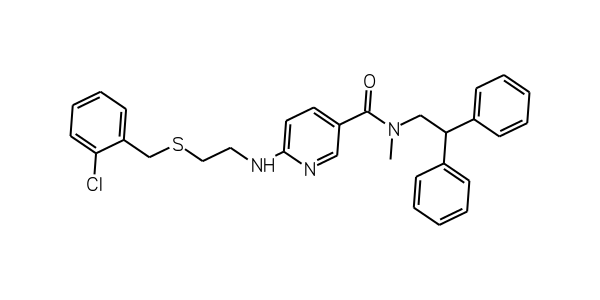} & {0.37 (0.5)} \\
ER$\alpha$ & \includegraphics[scale=0.6]{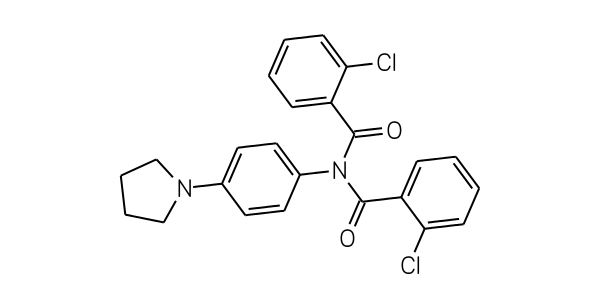} & 104 & \includegraphics[scale=0.6]{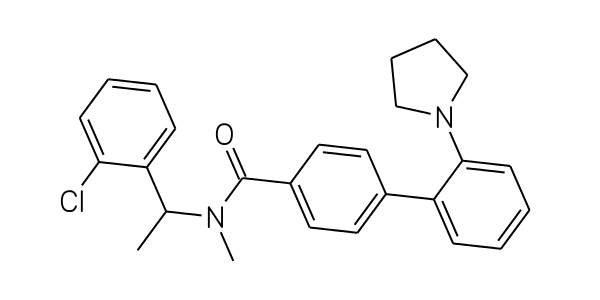} & {0.38 (0.52)} & \includegraphics[scale=0.6]{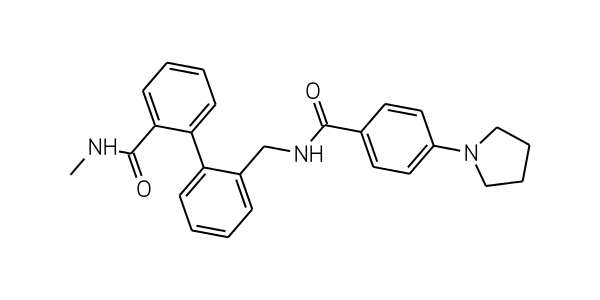} & {0.43 (0.51)} \\
ER$\alpha$ & \includegraphics[scale=0.6]{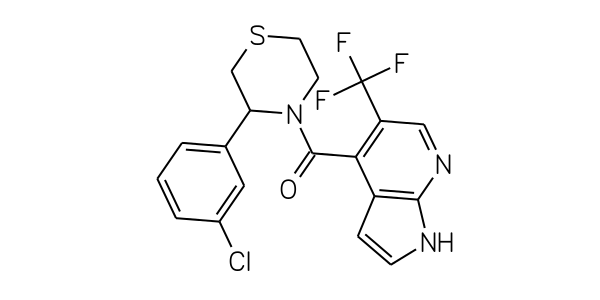} & 452 & \includegraphics[scale=0.6]{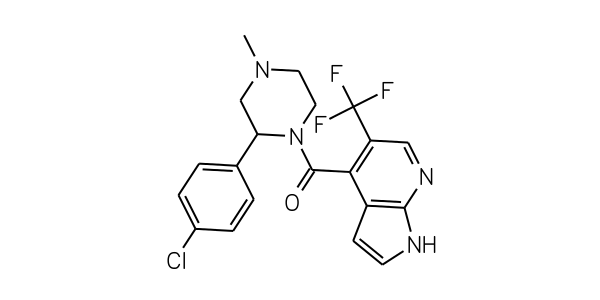} & {0.56 (0.63)} & \includegraphics[scale=0.6]{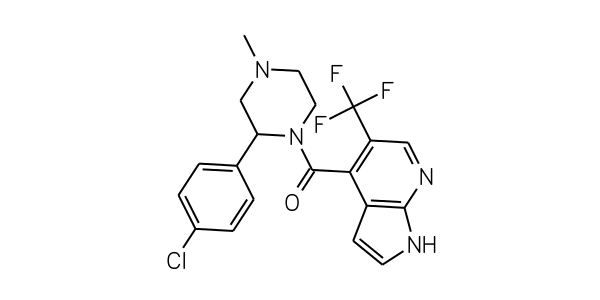} & {0.56 (0.63)} \\
c-KIT & \includegraphics[scale=0.6]{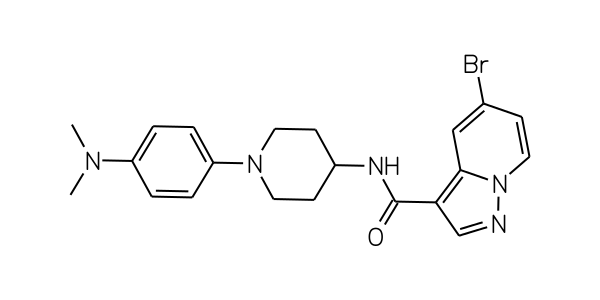} & 802 & \includegraphics[scale=0.6]{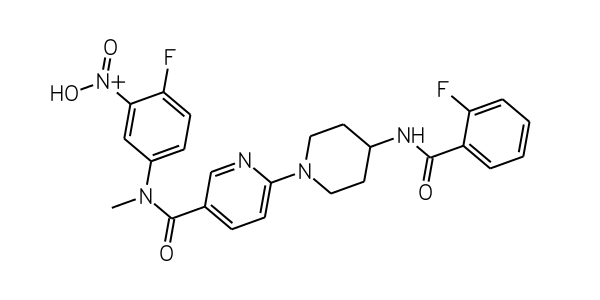} & {0.33 (0.34)} & \includegraphics[scale=0.6]{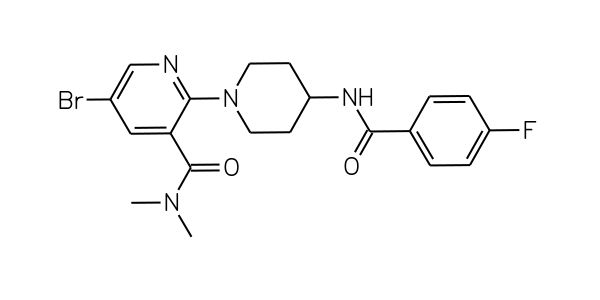} & {0.38 (0.39)} \\
c-KIT & \includegraphics[scale=0.6]{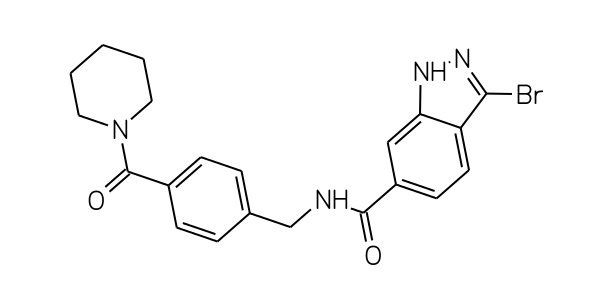} & 212 & \includegraphics[scale=0.6]{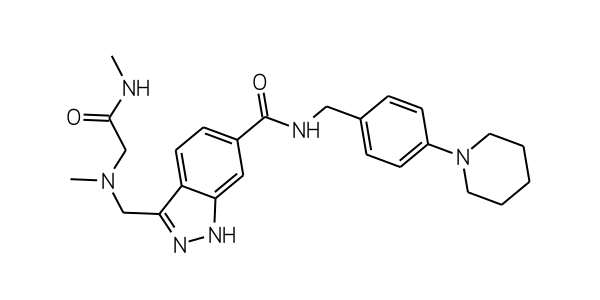} & {0.49 (0.44)} & \includegraphics[scale=0.6]{rdkit_structures/XLIB0016-__GL-040873_GL-032359.png} & {0.49 (0.44)} \\
c-KIT & \includegraphics[scale=0.6]{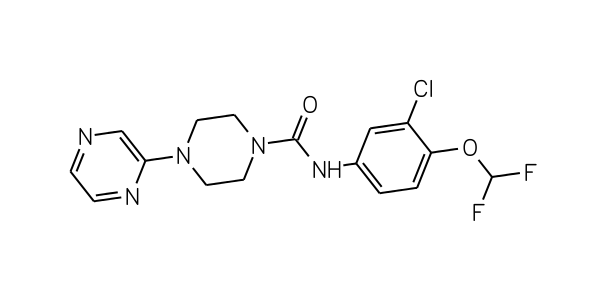} & 168 & \includegraphics[scale=0.6]{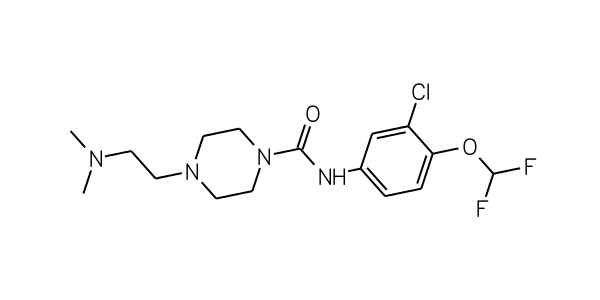} & {0.54 (0.47)} & \includegraphics[scale=0.6]{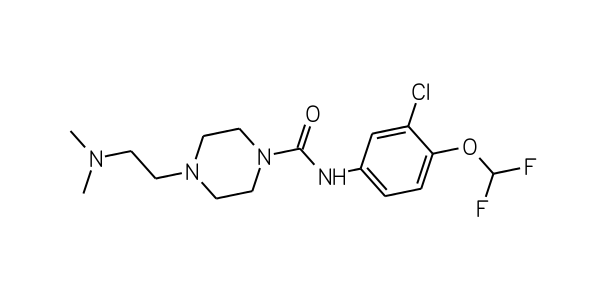} & {0.54 (0.47)} \\
c-KIT & \includegraphics[scale=0.6]{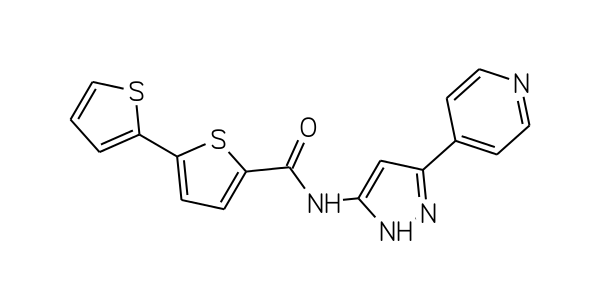} & 241 & \includegraphics[scale=0.6]{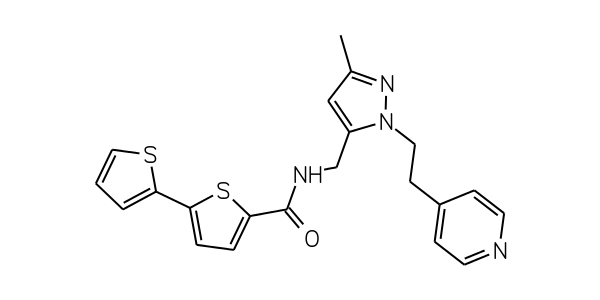} & {0.47 (0.47)} & \includegraphics[scale=0.6]{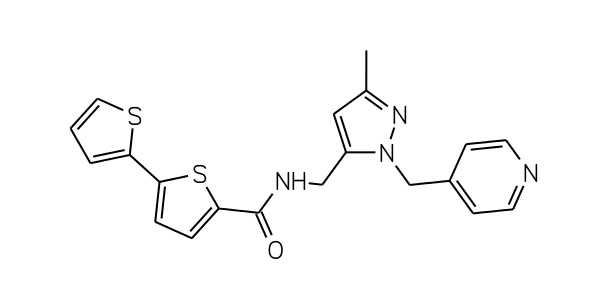} & {0.48 (0.48)} \\
\bottomrule
\end{tabular}
\caption{Nearest training set neighbors for confirmed GCNN hits selected to highlight model performance on compounds with low similarity to the training set. For each hit compound, we show the closest neighbors as measured by Tanimoto on ECFP6-counts fingerprints; it is possible that other near neighbors (including those from other similarity metrics) have structural features that are being integrated by the model. Similarity values are given as ECFP6-counts (FCFP6-counts).}
\label{table:hits}
\end{table}

\begin{table}[htb]
\centering
\sisetup{detect-weight=true,detect-inline-weight=math}
\rowcolors{2}{lightgray}{}
\begin{tabular}{c c c S S}
\toprule
{Target} & {Model} & {Source} & {\makecell{Concentration \\ ({\textmu}M)}} & {\makecell{Threshold \\ (\% inhibition)}} \\
\midrule
\cellcolor{white} & \cellcolor{white} & Mcule & 3 & 50 \\
\cellcolor{white} & \multirow{-2}{*}{\cellcolor{white} GCNN} & XVL & 1 & 50 \\
\cellcolor{white} & \cellcolor{white} & Mcule & 10 & 65 \\
\multirow{-4}{*}{\cellcolor{white} sEH} & \multirow{-2}{*}{\cellcolor{white} RF} & XVL & {-} & {-} \\
\midrule
\cellcolor{white} & \cellcolor{white} & Mcule & 30 & 70 \\
\cellcolor{white} & \multirow{-2}{*}{\cellcolor{white} GCNN} & XVL & 10 & 50 \\
\cellcolor{white} & \cellcolor{white} & Mcule & 10 & 50 \\
\multirow{-4}{*}{\cellcolor{white} ER$\alpha$} & \multirow{-2}{*}{\cellcolor{white} RF} & XVL & 10 & 45 \\
\midrule
\cellcolor{white} & \cellcolor{white} & Mcule & 30 & 50 \\
\cellcolor{white} & \multirow{-2}{*}{\cellcolor{white} GCNN} & XVL & 30 & 50 \\
\cellcolor{white} & \cellcolor{white} & Mcule & 30 & 50 \\
\multirow{-4}{*}{\cellcolor{white} c-KIT} & \multirow{-2}{*}{\cellcolor{white} RF} & XVL & 30 & 50 \\
\bottomrule
\end{tabular}
\caption{Thresholds used for retesting compounds with full dose--response curves. For each (target, model, source) combination, a compound was retested if the percent inhibition exceeded the threshold at the given concentration. For example, a proprietary virtual library (XVL) compound predicted by a GCNN model for sEH was retested if it achieved $\geq$~50\% inhibition of sEH at 1~{\textmu}M.}
\label{table:cutoffs}
\end{table}

\begin{figure}[t]
\centering
\includegraphics[width=\linewidth]{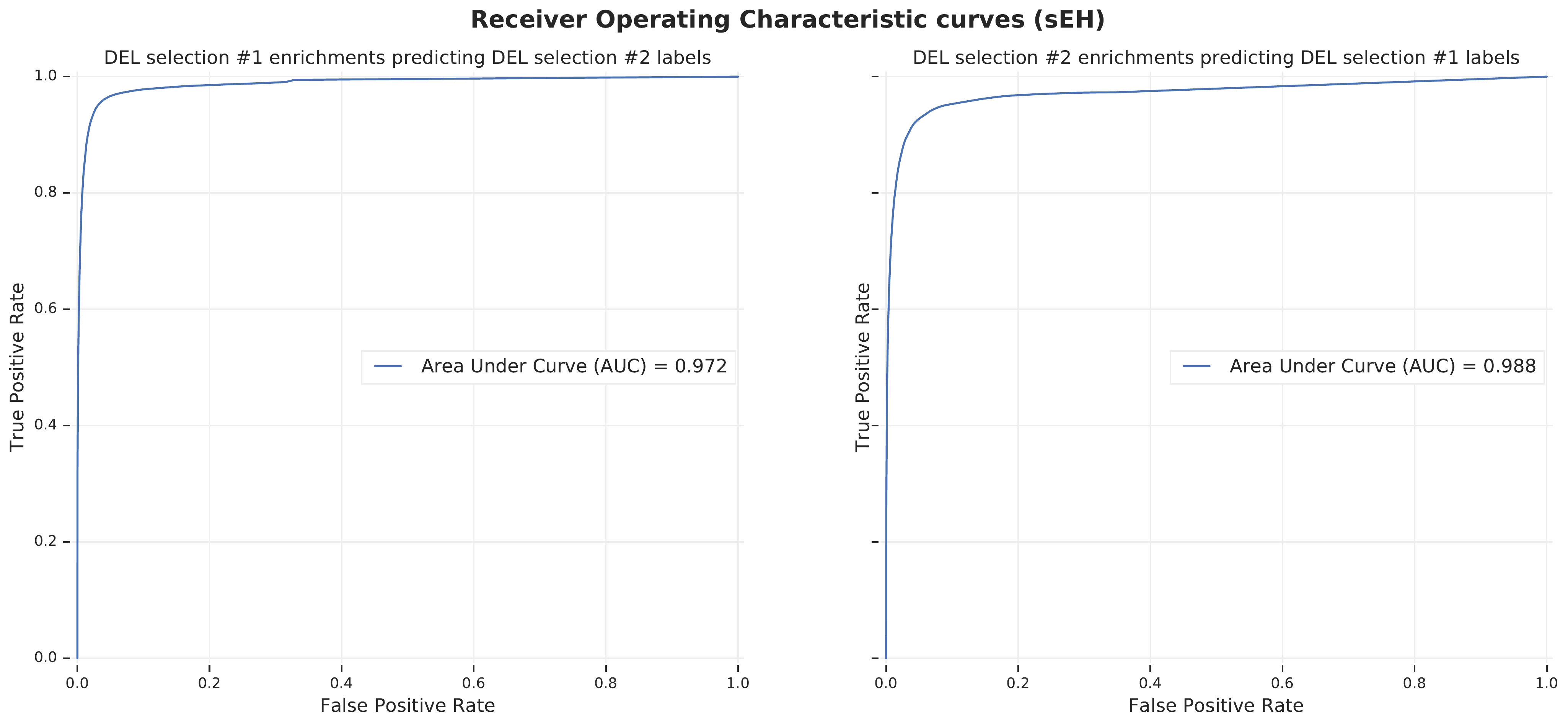}
\caption{Receiver Operating Characteristic curves of cross-prediction of two DEL selections on the sEH protein target, taken months apart, showing very good reproducibility.}
\label{fig:two_selections_roc_curve}
\end{figure}

\begin{table}[htb]
\centering
\sisetup{detect-weight=true,detect-inline-weight=math}
\begin{tabular}{l c c}
\toprule
 & {DEL} &     {DEL+ML}\\
\midrule
Computing cost for model building\textsuperscript{1} &         &  \$200 \\
Computing cost for inference\textsuperscript{2}      &         &  \$300 \\
\midrule
\textbf{Total computing cost}              &         &  \$500 \\
\midrule
Number of compounds acquired\textsuperscript{3}              & 75      &  370 \\
Synthesis cost per compound\textsuperscript{4}       & \$1,500 -- \$4,000  & \\
Acquisition cost per compound\textsuperscript{5}     &         &  \$50 -- \$200 \\
Assay cost per compound\textsuperscript{6}           & \$10     &  \$10 \\
\midrule
\textbf{Total compound cost}               & \$113,250 -- \$300,750 &  \$22,200 -- \$77,700 \\
\midrule
\textbf{Total cost}               & \$113,250 -- \$300,750 &  \$22,700 -- \$78,200 \\
\bottomrule
\end{tabular}
\caption{Approximate cost of the followup/validation of hits comparing a traditional DEL analysis with the ML approach described in this work. We left out the human analysis time in this process as it is difficult to make a fair estimate of the amount of time and the cost of a chemist and data scientist in the two approaches. As the table shows, the cost is primarily driven by the large difference in compound acquisition costs (custom synthesis is 20\textsf{x}-30\textsf{x} more expensive).
\\ \textsuperscript{1}~Derived from \$4.50/hr for a TPU on Google Compute Engine (\url{https://cloud.google.com/tpu/pricing}), each fold trains in 3 hours, and 15 folds per target (note that the number of folds varied across the targets in this study).
\\ \textsuperscript{2} Derived from \$0.01/hr on a CPU on Google Compute Engine (\url{https://cloud.google.com/compute/all-pricing}), 1M compounds per 30 CPU hours, and inference on 100M compounds.
\\ \textsuperscript{3} 75 is historically a normal number of compounds to synthesize off-DNA for X-Chem and 370 is the average number of compounds acquired for this study for the GCNN model. Note that given the hit rates achieved here, future applications of DEL+ML may not need as many compounds to produce useful molecules.
\\ \textsuperscript{4} Approximate range based on historical off-DNA synthesis costs paid by X-Chem.
\\ \textsuperscript{5} Range based on the price per compound from Mcule in this study.
\\ \textsuperscript{6} Approximate assay cost per compound in this study.
}
\label{table:del_vs_delml}
\end{table}

\end{document}